\documentclass[12pt]{article}

\usepackage{graphics}
\usepackage{epsfig}
\usepackage{graphicx}

\usepackage{times}

\newenvironment{sciabstract}{%
\begin{quote} \bf}
{\end{quote}}

\newcounter{lastnote}

\title{ 
``Explosive Percolation'' Transition 
\\ 
is Actually Continuous}

\author
{Rui A. da Costa,$^{1}$ Sergey N. Dorogovtsev,$^{1,2\ast}$ Alexander V. Goltsev,$^{1,2}$ \\
Jos\'e Fernando F. Mendes$^{1}$\\
\\
\normalsize{$^{1}$Departamento de F{\'\i}sica da Universidade de Aveiro, 3810-193 Aveiro, Portugal}\\
\normalsize{$^{2}$A. F. Ioffe Physico-Technical Institute, 194021
  St. Petersburg, Russia}\\
\\
\normalsize{$^\ast$To whom correspondence should be addressed; E-mail:  sdorogov@ua.pt.}
}

\date{}

\begin{document} 


\baselineskip24pt


\maketitle


\begin{sciabstract}
The basic notion of percolation in physics  
%%for disordered systems 
assumes the emergence of a giant connected (percolation) cluster 
in a large disordered system 
when the density of connections 
%%in a system 
exceeds some critical value. 
%%(percolation threshold).  
%%In 
%%%%classical 
%%%%the 
%%%%classical 
%%various percolation problems,  
%%%%, which are basic for understanding of disordered systems, 
%%increasing the number of 
%%%%random links 
%%connections between nodes of a network above some threshold 
%%%%leads to emergence of 
%%results in the emergence of
%%%%creates 
%%a giant connected (``percolation'') cluster. 
%%%%in this system. 
Until recently, the percolation phase transitions were believed 
%%known 
to be continuous, however, in 2009, a 
%%strikingly 
remarkably 
different, 
%%abrupt 
discontinuous phase transition was reported in a new so-called ``explosive percolation'' problem. 
Each link in this problem is 
%%chosen of a few possible ones 
established by a 
%%local 
specific optimization process. 
%%based on the sizes of clusters they connect. 
%%using some optimization rule. 
%%Merging clusters in this problem .... were selected to be minimal 
%%In the classical percolation problems, ..... transition is continuous... 
%%The understanding of disordered systems is essentially based on the classical percolation problem. 
%%In the 
%%classical 
%%the classical basic 
%%well known 
%%studied 
%%percolation problems  
%%Increase of 
%%increasing the number of random links between vertices above some threshold 
%%leads to emergence of 
%%creates a giant connected cluster in this system. This percolation transition is continuous, but 
%%in 2009, 
%%recently a strikingly different, abrupt transition was reported in a new ``explosive percolation'' problem, in which merging clusters 
%%for merging 
%%were selected of few ones in the framework of some minimization. 
Here,  
%%by using a representative model capturing 
%%%%which captures 
%%the essence of this kind of percolation, 
%%irreversible processes 
employing strict analytical arguments and 
%%extensive simulations, 
numerical calculations,  
we 
%%show 
find that in fact the ``explosive percolation'' transition is continuous though with an uniquely small critical exponent of the percolation cluster size. 
These transitions provide a new class of critical phenomena in irreversible systems and processes. 
\end{sciabstract}

% In setting up this template for *Science* papers, we've used both
% the \section* command and the \paragraph* command for topical
% divisions.  Which you use will of course depend on the type of paper
% you're writing.  Review Articles tend to have displayed headings, for
% which \section* is more appropriate; Research Articles, when they have
% formal topical divisions at all, tend to signal them with bold text
% that runs into the paragraph, for which \paragraph* is the right
% choice.  Either way, use the asterisk (*) modifier, as shown, to

\newpage

The modern understanding of disordered systems in statistical and condensed matter physics is essentially based on the notion of percolation 
%%problems 
({\it 1\/}). 
%%If we 
When one increases progressively the number of 
%%random 
%%links 
connections between nodes in a network, 
%%then 
%%, first the network consists of a set of finite components, but 
above 
%%at 
some critical number (percolation threshold) a giant connected (percolation) cluster 
%%is formed 
emerges 
in addition to finite clusters. This percolation cluster contains a finite fraction of nodes and links in a network. The percolation transition was 
%%well known 
widely believed 
to be a typical continuous phase transition 
%%with a critical power-law size distribution of finite clusters 
for various networks architectures and space dimensionalities ({\it 2\/}),  
so it 
%%demonstrates 
shows standard scaling features, including a power-law size distribution of finite cluster sizes at the percolation threshold. 
%%%%In particular, in the classical percolation problem for the infinite-dimensional systems, where mean-field theories are valid, the fraction of nodes belonging to the giant cluster (the order parameter) grows from zero proportionally to the excess of the number of connections over the percolation threshold value. 
%%This means that the corresponding critical exponent $\beta$ of the order parameter is $1$. 
Recently, however, it was reported that a remarkable percolation problem exists  
%%with a quite different, discontinuous 
%%%%, abrupt transition, 
in which the percolation cluster emerges 
%%abruptly 
discontinuously and already contains a finite fraction of nodes at the percolation threshold ({\it 3\/}). 
%%The percolation cluster in this problem emerges abruptly, with a jump.  
This conclusion was based on numerical simulations of a model in which each new connection is made in the following way: 
%%add at random two links to the network, and choose 
choose at random two links that could be added to the network, but add 
only one of them, namely the link connecting two clusters with the smallest product of 
%%cluster 
their sizes.  
%%not immediately between two randomly chosen nodes, but rather 
%%after two attempts to add a new link 
%%by choosing the link 
%%, by choosing one of the two potential connections, say, the link connecting two clusters with the smallest product of cluster sizes. 
%%It is just to 
To emphasize this surprising discontinuity, this 
%%type 
kind of percolation 
%%has been 
was named ``explosive'' ({\it 3\/}).   
Further investigations of ``explosive percolation'' in this and similar systems, 
%%(this term has been proposed to emphasize the abrupt emergence of ), 
also mainly based on numerical simulations, supported this strong result but, in addition,  
%%unexpectedly 
surprisingly for abrupt transitions, revealed power-law critical distributions of cluster sizes ({\it 4--10\/}) resembling those found in continuous percolation transitions. This self-contradicting combination of discontinuity and scaling have made explosive percolation one of the 
%%intriguing 
challenging and urgent issues in the physics of disordered systems. 

Here we resolve this confusion. 
We show that 
%%in fact 
there is 
%%actually no 
not actually any discontinuity at the ``explosive percolation'' threshold,  
%%discontinuity, and the explosive percolation transition is continuous 
%%in stark contrast 
contrary to the conclusions of the previous investigators.   
We consider a simple 
%%non-trivial, 
representative model demonstrating 
%%showing ``explosive percolation'' 
this new kind of percolation 
%%. 
%%with ultimate clarity. 
%%This 
%%%%infinite-dimensional model 
%%is 
%%%%quite 
%%%%very 
%%convenient for 
%%%%extensive 
%%numerical simulations (we simulate large systems of $2\times10^{9}$ nodes), it allows numerical solution of very large number of evolution equations for cluster size distributions (we solve equations for cluster sizes up to $10^6$ nodes in infinite systems), and, moreover, it allows strict analytical treatment. 
%%By using this model, we 
and show that the ``explosive percolation'' transition is 
a 
continuous, 
%%in contrast to what was believed previously. 
%%We find that in fact this is the 
second-order phase transition but, importantly, 
%%with an uniquely small critical exponent $\beta$ of the order parameter, 
with an uniquely small critical exponent $\beta\approx 0.0555$ of the percolation cluster size. 
%%so it is 
%%This is why it is very hard to distinguish this transition from a discontinuous one by 
%%%%applying 
%%numerical simulations, 
%%for 
%%even employing large system sizes. 
%%Furthermore, we show that even apart from the anomalously small critical exponent value, this unusual phase transition, though being continuous,  
%%%%this unusual transition 
%%remarkably differs from the classical percolation one.       
  
One of the simplest 
%%infinite dimensional 
systems in which classical percolation takes place is as follows. Start with $N$ unconnected nodes, where $N$ is large, and at each step add connection between two uniformly randomly chosen nodes. In essence, this is 
%%an 
a simple aggregation process ({\it 11\/}), in which at each step, a pair of 
%%selected 
clusters, to which these nodes belong, merge together 
%%, selected in this way, 
%%subsequently merge with each other 
(Fig.~1A). 
%%The probability that two clusters merge together is proportional to 
%%the 
%%product of the 
%%their sizes.   
%%It is convenient to 
If we introduce ``time'' $t$ as the ratio of the total number of added links in this system, $L$, and its size $N$, i.e., $t=L/N$, then the percolation cluster of relative size $S$ emerges at the percolation threshold $t_c=1/2$ and grows with $t$ in the following way: $S \sim \delta^\beta$, where $\delta=|t-t_c|$ and  $\beta=1$. 
%%, and grows with $t$.  
At the critical point, the cluster size distribution 
%%(concentration 
(fraction of finite connected components of $s$ nodes), $n(s)$, in this classical problem is power-law: $n(s) \sim s^{-\tau}$ with exponent $\tau=5/2$, see Ref.~({\it 1\/}). 
%%When the number $L$ of links in this network exceeds  
%%The classical percolation problem may be defined in the following way. 

In this report, we consider a direct generalization of this process. 
%%(Fig.~1B). 
Namely, at each step, we 
%%do two random samplings: 
%%take two samples: 
sample twice: 
%%, selecting two pairs of nodes: 
%%(i) choose two pairs of nodes uniformly at random, (ii) of each of these two pairs, select that node which belongs to the smaller 
%%
\begin{enumerate} 

\item[(i)] 
choose two nodes uniformly at random and compare the clusters to which these nodes belong; select that node of the two ones, which belongs to the smallest cluster; 

\item[(ii)] 
choose a second pair of nodes and, again, as in (i), select the node belonging to the smallest of the two clusters; 

\item[(iii)] 
add a link between the two selected nodes thus merging the two smallest clusters (Fig.~1B). 

\end{enumerate}
Repeat this procedure again and again. 
%%In principle
Note that in (i) and (ii) 
%%this scheme, 
%%One should note that 
%%the same 
a cluster can be selected several times. This 
%%%%is 
%%frequently 
%%%%the case 
%%occurs in the presence of 
is 
%%frequently 
the case for the percolation cluster. 
%%One should stress that this process is irreversible in contrast to ordinary percolation. 
%%In this special aggregation process, 
These rules contain the key element of other explosive percolation models, e.g., model ({\it 3\/}), namely, for merging, select the minimal clusters 
%%out of 
from a few possibilities. 
Importantly, our procedure provides even more 
%%effective 
stringent selection of small components for merging than model ({\it 3\/}) since guarantees that the product of the sizes of two merging clusters is the smallest of the four possibilities (each of the first pair of chosen nodes (i) may connect with any node of the second pair (ii)) in contrast to selection from only two possibilities in model ({\it 3\/}). Consequently, if we show that the transition in our model is continuous, than model ({\it 3\/}) also must have a continuous transition.  
%%the transition is continuous in our model, then it will be the case also for     
%%One should stress that this and other explosive percolation processes are irreversible in stark contrast to ordinary percolation, in which instead of adding connections one can equally eliminate them. 
%%---in the most strict and emphatic form. 
%%One should note that the same cluster can be selected several times. 
%%So, based on the previous studies, one could expect discontinuity  
More generally, 
%%in these random samplings, we can set the size of each sample to be $m$, 
in each sampling, one can select at random $m$ nodes 
%%we can set the 
%%%%size of each sample 
%%sample size in these random samplings to 
%%%%be 
%%$m$, 
thus choosing  
%%we can choose 
the minimal cluster 
%%not of $2$ but 
from $m$ clusters. 
%%, where $m$ is a positive integer number. 
Classical percolation corresponds to $m=1$ (Fig.~1A). 
%%For $m=1$, we have the classical percolation (Fig.~1A). 
In this report we mostly focus on $m=2$ (Fig.~1B). 
%%One should note that explosive percolation processes are irreversible in contrast to ordinary percolation.     
%%This specific aggregation process contains the main ingredient of other 
%%models of 
%%the existing models showing 
%%explosive percolation problems---for merging, the smallest clusters of a few ones are chosen. 
%%In our model, this selection is 
%%Moreover, this ... is present in our model in even more strong form than in or 
%%We can ...   stronger ....
%%Namely, ... Moreover .... in even stronger form .... 
%%This is a specific aggregation process, in which at each step we merge two minimal clusters chosen from the two pairs of clusters. This 
%%, which 
%%is similar to what ...... but even   stronger stronger form effect .... 
%%of m   --- of m  ...  the main ingredient --- select for merging the smallest clusters of a group of clusters is present here in even stronger form than in the original paper (4), so it should result in discontinuity.  belongs to the same class of models as ...  May select the same cluster several times      
One should stress that 
%%this and other 
the 
explosive percolation processes are 
irreversible in stark contrast to ordinary percolation. 
%%, in which instead of adding connections one can equally eliminate them. 
In the latter, one can reach any state 
%%is reachable 
either adding or removing connections. For explosive percolation, only adding links makes sense, and an inverse process is impossible. 
%%instead of adding, one can remove links at random, one by one. In  this way we can reach the same state. 
%%
%%as by adding connections.    

%%Thanks to simplicity, this model allows numerical simulations of unprecedentedly (for these problems) large systems. 
We simulated 
this irreversible aggregation process for a large system of 
%%an unprecedentedly (for these problems) large  
%%large 
%%network of 
$2{\times} 10^9$ nodes. 
%%($1000$ runs). 
%%At first sight (Fig.~1C), the 
When plotted over the full time range, 
the obtained 
dependence 
%%curve 
$S(t)$ 
%%when plotted over the 
shows what seems to be discontinuity at the critical point $t_c$ (Fig.~1C) similar to previous results, 
%%(Fig.~1C), 
but 
%%, but 
%%if we inspect the 
%%
a more thorough inspection of 
the critical region (log-log plot in Fig.~1D)
%%, however,  
%%more thouroghly, we  
%%shows 
reveals that 
%%the obtained data 
%%%%resulting curve 
%%does not 
%%%%allow us to 
%%exclude a continuous transition. 
%%%%one to claim that discontinuity exist or not. 
%%Indeed, Fig.~1D shows that 
%%Figure~1D 
%%only 
%%shows that the dependence $S(t)$ 
%%the resulting curve 
%%can be fitted by the function 
%%has a power-law contribution 
%%and 
%%can be  
the obtained data 
is 
definitely 
better fitted by the law $a\delta^\beta$, 
%%$S=a\delta^\beta$, 
which indicates a continuous transition, than, say,  
by $0.3+b\delta^\beta$. 
%%$S=0.3+b\delta^\beta$. 
%%laws. 
%%, and so, 
%%based on 
%%%%this data, we can only 
%%our simulations, we conclude that $0\leq S(t_c)<0.3$.    
%%but the data can be equally well fitted by the $S=a\delta^\beta$ and, say, $S=0.3+b\delta^\beta$ laws.   
%%%we cannot .....  .... fitting 
%%$S = a+b\delta^\beta$, where $a$ and $b$ are some coefficients, but values $a=0$ and, say, $a=0.5$ are both possible. 
%%Furthermore, even a more thorough finite-size analyse has not allow us to exclude a continuous transition.  
%%turn out to be both reasonable.  
%%NonethelFigure~1D, however, shows that     
%%with any $a$ within the interval $[0,0.5]$. That is, even that large size ($2\times 10^9$ nodes) is still not sufficient for a definite conclusion about the presence of discontinuity.   
%%, which are by many orders of magnitude larger than in the other studies of explosive percolation. 
%%Consequently, to resolve this 
%%For a definite conclusion, 
%%%%Thus, to resolve 
%%%%%%our 
%%%%this problem, 
%%%%it is necessary to study 
%%we must investigate 
%%%%have to study an infinite system. 
%%the infinite size limit using analytical arguments. 
Fitting the data of our simulation by the law $S_0+b\delta^\beta$, we find that 
%%at the critical point, 
$S_0$ is at least smaller than $0.05$. 
%%For a more 
%%So 
This shows that for a definite conclusion, even so large a system turns out to be not sufficient, and 
%%Nonetheless, this plot demonstrates that even so large system size is not sufficient, and  
a discontinuity can be ruled out or validated only by analytical arguments for the infinite size limit.    
%%A 
%%%more definite 
%firm conclusion must be based on analytical arguments for the infinite size limit.   
%%There are still two possibilities, and one of them can be ruled out only by analytical arguments for the infinite size limit.   

%%As is usual for percolation, we 
We address this problem analytically and numerically by considering    
%%let us consider 
the evolution of the size distribution $P(s)$ for a finite cluster of $s$ nodes to which a randomly chosen node belongs: $P(s) = sn(s)/\langle s \rangle$, 
%%$\sum_s P(s)=1-S$, 
%%. Here the mean cluster size is 
where $\langle s \rangle$ 
%%$n(s)$ is the concentration of clusters of size $s$ and 
%%$\langle s \rangle=\sum_s s n(s)$ is the mean cluster size.  
is the average cluster size (the ratio of the number of nodes and the total number of clusters). 
%%$\langle s \rangle$ is the average cluster size (including the percolation cluster).  
This distribution satisfies the sum rule $\sum_s P(s)=1-S$. 
%%In this model, however, the distribution of merging clusters 
%%In addition, 
Another important characteristic in this model is the probability $Q(s)$ that if we choose at random two nodes than the smallest of the two clusters to which these nodes belong is of size $s$. 
$Q(s)$ provides us with the size distribution of merging clusters.    
%%Importantly, in this model, the size distribution $Q(s)$ of merging clusters significantly differs from $P(s)$. 
Here $\sum_s Q(s)=1-S^2$. 
%%$Q(s)$
If we introduce the cumulative distributions 
$P_{\rm cum}(s) \equiv \sum_{u=s}^\infty P(u)$ and $Q_{\rm cum} \equiv \sum_{u=s}^\infty Q(u)$, then probability theory gives $Q_{\rm cum}(s)+S^2 = [P_{\rm cum}(s)+S]^2$. 
%%, which is a known result from probability theory. 
%%That is, 
Consequently  
\begin{equation}
Q(s) = [P_{\rm cum}(s)
%%+S
+P_{\rm cum}(s+1)+2S]P(s) 
= [2-2P(1)
%%-2P(2)
-\ldots-2P(s-1)-P(s)]P(s)
, 
\label{e10}
\end{equation}
%%
%%which shows that 
that is $Q(s)$ is determined by $P(s')$ with $s'\leq s$.
%%One can show that the 
The evolution of these distributions in the infinite system is 
%%explicitly 
exactly described by the master equation: 
%%the following generalization of the standard Smoluchowski equation: 
%%
\begin{equation}
\frac{\partial P(s,t)}{\partial t}
= s \sum_{u+v=s} Q(u,t)Q(v,t) -2 sQ(s,t)
, 
\label{e20}
\end{equation}
%%
%%where the 
which generalizes the standard Smoluchowski equation. 
We derived Eq.~(2) only assuming the infinite system size.   
The only difference from the classical percolation problem ({\it 11\/}) is the presence of the distribution $Q(s,t)$ instead of $P(s,t)$ 
%%for classical percolation 
on the right-hand side of 
this
%%the corresponding 
equation.  
Thus we have a chain of coupled equations, which should be solved analytically or numerically. To find a numerical solution, first solve the first equation of the chain, which gives $P(1,t)$. Substitute this result into the second equation and solve it, which gives $P(2,t)$, and so on. 
%%$P_{\rm cum} \equiv \sum_{u=s}^\infty P(s)$ 
In this way we 
%%These equations allow us to 
find numerically the distributions $P(s,t)$ and $Q(s,t)$ at any $t$ for infinite $N$. Solving $10^6$ equations gives 
%%, we have obtained the complete description of 
the evolution of these distributions for $1{\leq} s {\leq} 10^6$ and $S(t) 
%%\approx 
\cong 1-\sum_{s=1}^{10^6}P(s,t)$. 
%%This wide range 
%%is already sufficient for 
%%allows us to arrive at solid conclusions. 
%%
The 
%%resulting 
log-log plot 
%%for $S$ vs. $\delta=t-t_c$ 
(Fig.~2A) shows that the obtained $S(t)$ dependence is well described by the power law: $S \propto \delta^\beta$ with $\delta=t-t_c$, $t_c=0.923207508(2)$, and small $\beta=0.0555(1)$ (which is very close if not equal to $1/18=0.05555...$). 
Here we find $t_c$ as the point at which $P(s)$ 
%%and $Q(s)$ are 
is power-law over the full range of $s$, see below. 
To check the correctness and precision of our calculations, we repeated them
for ordinary percolation and obtained the classical results with the same
presicion as for our model. 
%%Moreover, although 
Although the small exponent $\beta$ makes it difficult to approach the narrow 
%%critical 
region of small $S$, fitting this data by the law $S_0+b\delta^\beta$, we find that 
%%at the critical point, 
$S_0$ is smaller than $0.005$. This supports our hypothesis that the transition is continuous, 
but still does not prove it. 
Moreover, both our extensive simulations and numerical solution results clearly demonstrate that
the analysis of the $S(t)$ data cannot validate or rule out a discontinuity. 
Figure~2B shows the evolution of the distribution $P(s,t)$, which we  
%%in the full time range. 
%%Compare this evolution 
compare with the evolution of this distribution for ordinary percolation, Fig.~2C. The difference is strong at $t<t_c$, where the distribution for explosive percolation has a bump, but above $t_c$ the behaviors are similar.    
%%(compare it to the evolution of this distribution for ordinary percolation, Fig.~2C and note a strong difference at $t<t_c$, where 
%%%%$P(s)$ 
%%the distribution for explosive percolation has a peak). 
The distribution function $Q(s,t)$ evolves similarly to $P(s,t)$ in the full time range.  
%% and also has a peaked form at $t<t_c$.  
%%of times. 
At the critical point, we find power-law $P(s) \sim s^{1-\tau}$ and $Q(s) \sim s^{3-2\tau}$ in the full range of $s$ (six orders of magnitude), where $\tau=2.04762(2)$, which is close to $2$, as in ({\it 7,8,10\/}),  
%%(compare with 
in contrast to $\tau=5/2$ for classical percolation. 
%%The obtained $\tau$ is within the range of values found for other explosive percolation problems ({\it 5--15, 18\/}). 
%%). 
%%We also find that the 
The first moments of these distributions, $\langle s \rangle_{\scriptscriptstyle \!P} \equiv \sum_s s P(s)$ 
(the mean size of a finite cluster to which a randomly chosen node belong)  
and $\langle s \rangle_{\scriptscriptstyle \!Q} \equiv \sum_s s Q(s)$, demonstrate power-law critical singularities  
%%singularities, 
$\langle s \rangle_{\scriptscriptstyle P} \sim |\delta|^{-\gamma_P}$ and $\langle s \rangle_{\scriptscriptstyle Q} \sim |\delta|^{-\gamma_{\scriptscriptstyle Q}}$, where exponents $\gamma_{\scriptscriptstyle P}=1.0111(1)$ and $\gamma_{\scriptscriptstyle Q}=1.0556(5)$ both below and above the transition. Note that $\gamma_{\scriptscriptstyle P}>1$ in contrast to ordinary percolation, where the mean-field value of exponent $\gamma$ is $1$. 
Figure~2D shows the set of time dependencies of $P(s,t)$ for fixed cluster sizes (the time dependencies of $Q(s,t)$ are similar). 
%%Note that these 
These dependencies strongly differ from those for ordinary percolation (Fig.~2E) in the following respect.  
%%, which are symmetrical with respect to $t_c$ at large $s$. 
%%With increasing $s$, the 
%%series of 
The peaks in Fig.~2D for explosive percolation are 
%%displaced to the left-hand side of the plot, 
below $t_c$, 
%%from $t_c$, 
while the peaks in Fig.~2E for ordinary percolation are symmetrical with respect to the critical point 
%%$t_c$ 
at large $s$.  
%%Figures~2D and 2E show the set of time dependences $P(s,t)$ and $Q(s,t)$ 
%%%%vs. time $t$ 
%%for fixed cluster sizes. 
%%Note the unusual nonmonotonous, peaked form of these curves. 
%%Recall that for classical percolation, the corresponding curves are exponentially increasing with time. ... 
%%Note the nonmonotonous form of these curves. 

The 
%%analysis 
inspection of these numerical results 
%%dependences 
in the critical region ($t<t_c$) reveals 
%%confirms 
%%the 
a scaling behavior 
%%of these distributions 
typical for continuous phase transitions,  
%%and allows us to obtain two scaling functions, 
%%for example 
%%the scaling functions $f(x)$ and $g(x)$ for $P(s,t) = s^{1-\tau}f(s\delta^{1/\sigma})$ and $Q(s,t) = s^{3-2\tau}g(s\delta^{1/\sigma})$, respectively, where $\sigma=(\tau-2)/\beta$, which is a standard scaling relation. 
%%This is a typical scaling behavior 
$P(s,t) = s^{1-\tau}f(s\delta^{1/\sigma})$ and $Q(s,t) = s^{3-2\tau}g(s\delta^{1/\sigma})$, respectively, where $f(x)$ and $g(x)$ are scaling functions, and $\sigma=(\tau-2)/\beta$, which is a standard scaling relation. One should stress that these functions are quite unusual.  
%%What is unusual is a form of these scaling functions. 
%%We obtain 
%%Therefore $\sigma=0.375$. 
%%Figure~3A 
Figure~3 shows the resulting scaling functions and, for comparison, the scaling function for ordinary percolation.  
%%To confirm that these are indeed scaling functions, we repeated our calculations with other initial conditions, e.g., with power-law distributed clusters at the initial instant, and obtained the same $f(x)$ and $g(x)$. 
Remarkably, $f(x)$ and, especially, $g(x)$ are well fitted by Gaussian functions. 
%%Note that these 
These 
%%peaked 
functions differ dramatically from the monotonously 
%%(exponentially) 
decaying exact scaling function $e^{-x}/\sqrt{2x}$ for ordinary percolation. 
%%(Fig.~3B). 
%%in two respects. 
Effective 
%%disappearance 
elimination of the smallest clusters in this merging process results in the
minima of the scaling functions at $x{=}0$. On the other hand, the stunted
emergence of large clusters results in the particularly rapid decay of these
functions at $x\gg1$.  
%%(i) Effective elimination of 
%%%%smallest 
%%minimal clusters in this evolution process leads to the rapid (Gaussian) decay of the scaling functions. (ii) In the normal phase, small clusters mostly merge with other small clusters producing again small clusters, which results in the peaked form of the scaling functions. In other words, the emergence of large clusters is suppressed. 

The key point of our report is the following strict analytical derivation. We
start from the fact that in this problem the cluster size distributions are power-law at the 
%%critical point 
percolation threshold. 
%%are power-law.  
%%observed 
We observed these power-laws over $6$ orders of magnitude, and they were
observed in works ({\it 7,8,10\/}) for explosive percolation though in less wide range of $s$. 
Now 
%%let us 
we strictly show that if the cluster size distribution is power-law at the
critical point, then this phase transition is continuous. 
%%The power-law cluster size distributions at the 
%%%%critical point 
%%percolation threshold, which we reliably observe 
%%%%observed 
%%in a so wide range of $s$, are of primary importance for this problem. 
%%(Note 
%%%%One should stress 
%%that it is much easier to observe the scale-free distribution in this problem than to measure the $S(t)$ in the critical region.)  
%%We use this feature to 
%%%%Based on this property 
%%%%%%This property 
%%%%%%%%actually 
%%%%%%guarantees 
%%%%we 
%%%%can 
%%%%strictly derive 
%%derive strictly 
%%the continuity of the phase transition.  
%%%%The point is 
We use the fact that in the critical region above the percolation threshold, 
%%the large $s$ asymptotics of the 
%%at large $s$ 
the distributions $Q(s)$ and $P(s)$ at large $s$ are proportional to each other, namely $Q(s) \cong 2S P(s)$, see Eq.~(1). 
This relation crucially simplifies our problem, since the resulting evolution
equation for the asymptotics of the distribution in this region contains only
$P(s)$ and the relative size $S(t)$ of the percolation cluster. 
%%This equation 
As a result, Eq.~(2) 
becomes 
very 
similar to that for ordinary percolation (the only difference is the presence
of $S(t)$ terms on the right-hand side), and so it can be easily analysed 
explicitly 
%%by using standard methods developed for percolation and
%%aggregation processes 
in the same way as for ordinary percolation ({\it 11}).   
%%One can show that to obtain the size of the critical 
%%Due to this 
%%%%relation, 
%%simplification, the evolution equations in this region 
%%%%become 
%%are very similar to those for standard percolation and can be easily analysed explicitly. 
%%This simplification allows us to analyse Eq.~(2) explicitly in this region.    
%%%%Based on the power-law form of the size distribution of clusters at the ....... 
%%In the critical region above the percolation threshold, where $Q(s) \cong 2S P(s)$ at large $s$, Eq.~(2) can be 
%%%%easily 
%%analysed explicitly. 
%%We use 
In this way, using a power-law critical distribution $P(s) \sim s^{1-\tau}$ 
%%, according to our numerical findings, 
%%which we obtained numerically, 
as an initial condition for 
%%these equations 
this equation at $t=t_c$, we find that the behavior of the distribution $P(s)$ and $S(t)$ in explosive percolation above 
%%$t_c$ 
the percolation threshold is qualitatively similar to that for ordinary percolation.  
%%we 
%%have proved 
%%prove 
%%one can 
Specifically, we show that the percolation cluster emerges continuously, 
and 
$S \propto \delta^\beta$, where $\tau = 2 + \beta/(1+3\beta)$ and so $\sigma=1/(1+3\beta)$. 
%%In other words, we 
%%%%find 
%%demonstrate that a scale-free form of the distributions at the critical point inevitably leads to a continuous phase transition in these problems. 
%%One can see that 
The obtained numerical values of exponents $\tau$, $\beta$, and $\sigma$ 
%%(Table 1) 
%%obtained by numerical solution of the evolution equations satisfy 
agree with these 
two 
scaling relations and thus also confirm the continuous transition. 
So the results of this report are self-consistent. 
Our results are summarized in Table~1.  
Assuming a scaling form for the distributions gives 
%%$\gamma_{\scriptscriptstyle \!P}=(3-\tau)/\sigma$ and $\gamma_{\scriptscriptstyle \!Q}=(5-2\tau)/\sigma$, and so we have 
$\gamma_{\scriptscriptstyle \!P} = 1+2\beta$ and $\gamma_{\scriptscriptstyle \!Q} = 1+\beta$, which agree with our numerical solution of Eq.~(2). Furthermore, applying standard scaling relations ({\it 1\/}), we calculate the fractal dimension for this model, $d_f=2/\sigma=2(1+3\beta)
%%=2....
$, and the upper critical dimension, $d_u = d_f+2\beta = 2(1+4\beta)
%%=2....
$ (see Table 1). 
%%(see Table 1 for the complete set of numerical results and comparison to ordinary percolation). 
The latter determines the finite size effect: 
$t_c(\infty)- t_c(N)
%%\Delta t_c(N) 
\propto N^{-2/d_u}$, where $2/d_u=0.818(1)$. 
Interestingly, the obtained fractal and upper critical dimensions for explosive percolation are less than $3$. They are much smaller than those for ordinary percolation, which are $4$ and $6$, respectively. 
Our model is infinite-dimensional, that is above the upper critical dimension, 
%%in situation 
where mean-field theories must work, which makes the observed smallness of exponent $\beta$ particularly remarkable. 
%%surprising.   
We know no other model in statistical physics with 
%%such 
a small $\beta$ 
in such a situation. 
%%above an upper critical dimension.   

In the general case, each of the samplings in the process involves $m\geq 1$ nodes, and so the minimal cluster is selected from the $m$ possibilities. 
%%We find that 
%%One can make the selection of minimal clusters even more emphatic by increasing $m$ (the number of attempts of which we choose the minimal cluster). It turns out that even this increase does not leads to discontinuity. We have found 
In this case, the relation between between the distributions $Q(s)$ and $P(s)$ is substituted by a more general one, but the form of the evolution equation (2) does not change.  
%%This generalization changes the form of the relation between the distributions $Q(s)$ and $P(s)$, but the 
We find that with increasing $m$, $t_c$ approaches $1$ and $\beta \cong \tau-2 \sim m^{-1}e^{-m}$, that is, $\beta$ rapidly decreases with $m$, but the transition remains continuous. 
Thus the critical exponents depend on $m$, and so 
%%These results for exponents show that 
the ``explosive percolation'' transitions have no universal critical behavior. 
%%in terms of critical exponents, but that this transition is generally continuous. 
%%We emphasize that the choosing of the smallest clusters in our representative model is performed in a stronger form than in the original model ({\it 3\/}), so we can state the absence of discontinuity in the model ({\it 3\/}) as well as in the other processes of this kind.     

We have shown that the ``explosive percolation'' transition is actually 
continuous. It is the smallness of the $\beta$ exponent for the size of the percolation cluster that makes it virtually impossible to distinguish this phase transition from a discontinuous one even in very large systems. 
Indeed, suppose 
%%Suppose 
that $N=10^{18}$ and $\beta\approx 1/18$. 
%%The smallest time
The addition of one link changes $t$ by $\Delta t=1/N$, which is the 
smallest time interval in the problem. Then a single step $\Delta t=10^{-18}$ from the percolation threshold already gives $S \sim (\Delta t)^\beta \sim 0.1$. 
%%minimal observable change of $t$. 
%%Then the minimal measurable deviation from the percolation threshold by $10^{-9}$ already gives a huge jump of the order of $
%%%%S(t_c+10^{-9}) \sim 
%%10^{-1/2} \sim 1$. 
%%The combination of uniquely small exponents and unusual scaling functions results in .... 
Other critical exponents and dimensions also differ radically from classical values. 
Furthermore, we have derived a complete set of scaling relations between the critical exponents for this problem, which 
%%agree with 
were also supported by our numerical results. 
%%While these relations differ from those for ordinary percolation, they show that the transition is continuous.   
The real absence of 
%%discontinuity is a somewhat shocking result given it 
explosion 
topples an already 
%%well-
established view of explosive percolation. We believe, however, that, thanks to the observed unique properties of this phase transition, our findings make this 
new class of irreversible systems 
%%processes 
%%problems 
an even more appealing subject 
%%of 
for further extensive exploration.  

\bibliography{scibib}

\bibliographystyle{Science}

% Following is a new environment, {scilastnote}, that's defined in the
% preamble and that allows authors to add a reference at the end of the
% list that's not signaled in the text; such references are used in
% *Science* for acknowledgments of funding, help, etc.

%%\begin{scilastnote}
%%\item We've included in the template file \texttt{scifile.tex} a new
%%environment, \texttt{\{scilastnote\}}, that generates a numbered final
%%citation without a corresponding signal in the text.  This environment
%%can be used to generate a final numbered reference containing
%%acknowledgments, sources of funding, and the like, per {\it Science\/}
%%style.  Along those lines, we'd like to thank readers of this document
%%for their attention, and invite them to address any questions to
%%Stewart Wills, at swills@aaas.org.
%%\end{scilastnote}

\newpage

\begin{quote}
{\bf References and Notes}

\begin{enumerate}

\item
D.~Stauffer and A.~Aharony, 
{\em Introduction to percolation theory\/} (Taylor \& Francis, London, 1994).

\item
S.~N.~Dorogovtsev, A.~V.~Goltsev, and J. F. F. Mendes, 
%%Critical phenomena in complex networks, 
{\em Rev.\ Mod.\ Phys.} {\bf 80}, 1275 (2008).

\item
D.~Achlioptas, R.~M. D'Souza, and J.~Spencer, 
%%Explosive percolation in random networks, 
{\em Science\/} {\bf 323}, 1453 (2009).

\item
Y.~S.~Cho, J.~S.~Kim, J.~Park, B.~Kahng, and D.~Kim, 
%%Percolation Transitions in Scale-Free Networks under Achlioptas Process, 
{\em Phys.\ Rev.\ Lett.} {\bf 103}, 135702 (2009). 
%%arXiv:0907.0309

\item
F.~Radicchi and S.~Fortunato, 
%%Explosive percolation in scale-free networks, 
{\em Phys.\ Rev.\ Lett.} {\bf 103}, 168701 (2009). 
%%arXiv:0907.0755

\item
R.~M.~Ziff, 
%%Explosive growth in biased dynamic percolation on two-dimensional regular lattice networks,  
{\em Phys.\ Rev.\ Lett.}  {\bf 103}, 045701 (2009). 
%%arXiv:0903.4678 

%%\item
%%Y.~S.~Cho, B.~Kahng, and D.~Kim,  
%%Cluster aggregation model for discontinuous percolation transition, 
%%arXiv:0911.4001. 

\item
F.~Radicchi and S.~Fortunato,  
%%Explosive percolation: a numerical analysis, 
{\em Phys.\ Rev.} E {\bf 81}, 036110 (2010). 
%%arXiv:0911.3549 

\item
R.~M.~D'Souza and M.~Mitzenmacher, 
%%Local cluster aggregation models of explosive percolation, 
{\em Phys.\ Rev.\ Lett.}  {\bf 104}, 195702 (2010). 
%%arXiv:1001.5030. 

%%\item
%%E.~J.~Friedman and J.~Nishimura, 
%%Explosive Percolation in Social and Physical Networks, 
%%arXiv:1001.4772. 

\item
R.~M.~Ziff, 
%%Scaling behavior of explosive percolation on the square lattice, 
arXiv:0912.1060. 

%%\item
%%S.~S.~Manna and A.~Chatterjee, 
%%A new route to Explosive Percolation, 
%%arXiv:0911.4674. 

%%\item
%%H.~D.~Rozenfeld, L.~K.~Gallos, H.~A.~Makse, 
%%Explosive Percolation in the Human Protein Homology Network, 
%%arXiv:0911.4082. 

%%\item
%%A.~A.~Moreira, E.~A.~Oliveira, S.~D.~S.~Reis, H.~J.~Herrmann, and J.~S.~Andrade Jr., 
%%A Hamiltonian approach for explosive percolation, 
%%{\em Phys.\ Rev.} E {\bf 81}, 040101 (2010). 
%%arXiv:0910.5918

%%\item
%%E.~J.~Friedman and A.~S.~Landsberg, 
%%Construction and Analysis of Random Networks with Explosive Percolation, 
%%{\em Phys.\ Rev.\ Lett.} {\bf 103}, 255701 (2009) 
%%%%arXiv:0910.3979

%%\item
%%D.~Stauffer, 
%%Scaling theory of percolation clusters, 
%%{\em Phys.\ Rep.}  {\bf 54}, 1 (1979). 

\item
Y.~S.~Cho, S.-W.~Kim, J.~D.~Noh, B.~Kahng, and D.~Kim, 
%%Self-organization in explosive percolation transitions, 
arXiv:1006.2194. 

\item
F. Leyvraz, 
%%Scaling theory and exactly solved models in the kinetic of irreversible aggregation, 
{\em Phys.\ Rep.}  {\bf 383}, 95 (2003). 

%%\item
%%N.~A.~M.~Ara\'ujo and H.~J.~Herrmann, 
%%Controlling the largest cluster detonates explosive percolation, 
%%arXiv:1005.2504. 

%%\item
%%Y.~S.~Cho, S.-W.~Kim, J.~D.~Noh, B.~Kahng, and D.~Kim, 
%%Self-organization in explosive percolation transitions, 
%%arXiv:1006.2194. 

\item
We thank S. Fortunato and F. Radicchi for useful information and many stimulating discussions on explosive percolation and G.~J. Baxter for helpful comments. This work was partially supported by POCI projects No. FIS/108476/2008 and SAU-NEU/103904/2008, and by the ARTEMIS and SOCIALNETS EU projects.

%%\item
%%***********
%%
%%\item 
%%G. Gamow, {\it The Constitution of Atomic Nuclei and
%%Radioactivity\/} (Oxford Univ. Press, New York, 1931).
%%
%%\item 
%%W. Heisenberg and W. Pauli, {\it Zeitschr.\ f.\ Physik} {\bf 56},
%%1 (1929).
\end{enumerate}
\end{quote}

\begin{table}[h] \centering
\caption{
Threshold values, critical exponents, and fractal and upper critical dimensions for ordinary percolation and explosive ($m=2$) one. Critical exponents for explosive percolation are expressed in terms of exponent $\beta$: $\tau=1+\beta/(1+3\beta)$, $\sigma=1/(1+3\beta)$, $\gamma_{\scriptscriptstyle P}=(3-\tau)/\sigma=1+2\beta$, $\gamma_{\scriptscriptstyle Q}=(5-2\tau)/\sigma=1+\beta$, $d_f=2(1+3\beta)$, $d_u=2(1+4\beta)$\vspace{5pt}.
}
\begin{tabular}
%%[c]{l|c|c}
{lcccccccc}
\hline
&  
& 
&
&
& 
&
&
&
\\[-10pt]
& 
$t_c$  
& 
$\beta$ 
&
$\tau$
&
$\sigma$
&  
$\gamma_{\scriptscriptstyle P}$ 
&
$\gamma_{\scriptscriptstyle Q}$ 
&
$d_f$
&
$d_u$
\\[-10pt]
&  
& 
&
&
&  
&
&
&
\\
\hline
&  
& 
&
&
&  
&
&
&
\\[-6pt]
${}\!$Ordinary${}\!$ 
%%percolation 
& 
$1/2$ 
& 
$1$ 
&
$5/2$
&
$1/2$
& 
$1$
&
---
&
$4$
&
$6$
\\[-6pt]
&  
& 
&
&
&  
&
&
&
\\
\hline
&  
& 
&
&
&  
&
&
&
\\
[-6pt]
${}\!$Explosive${}\!$ 
%%percolation 
& 
$\!0.923207508(2)\!$  
& 
$\!0.0555(1)\!$ 
%%%%%%%%%%0.5555555555556=1/18
&
$\!2.04762(2)\!$
%%%%%%%%%%2.04762=2+1/21
&
$\!0.857(3)\!$
& 
$\!1.111(1)\!$ 
&
$\!1.0556(5)\!$
&
$\!2.333(1)\!$
&
$\!2.445(1)\!$
\\[-6pt]
&  
&  
&
&
& 
&
&
&
\\
\hline
%%&  & 
%%\\[-6pt]
%%Autocorrelator 
%%& 
%%$\exp$
%%& 
%%Eqs.$~$(
%%\\[-6pt]
%%&  & 
%%\\\hline
%%&  & 
%%\\[-6pt]%
%%Propagator at  
%%& 
%%$\left[\right]$ 
%%& 
%%$\left[ \frac{\left(  l\right)}{t}\right]$ 
%%\\[-6pt]
%%&  & 
%%\\
%%\hline
%%&  & 
%%\\[-6pt]
%%Propagator   
%%& 
%%$\mu$
%%& 
%%$t$
%%\\[-6pt]
%%&  
%%& 
%%\\\hline
\end{tabular}
\label{results}
\end{table}

$\phantom{.}$\newpage{}

\begin{figure}[t]
%%%%[tbhd]
\begin{center}
\scalebox{0.27}{\includegraphics[angle=0]{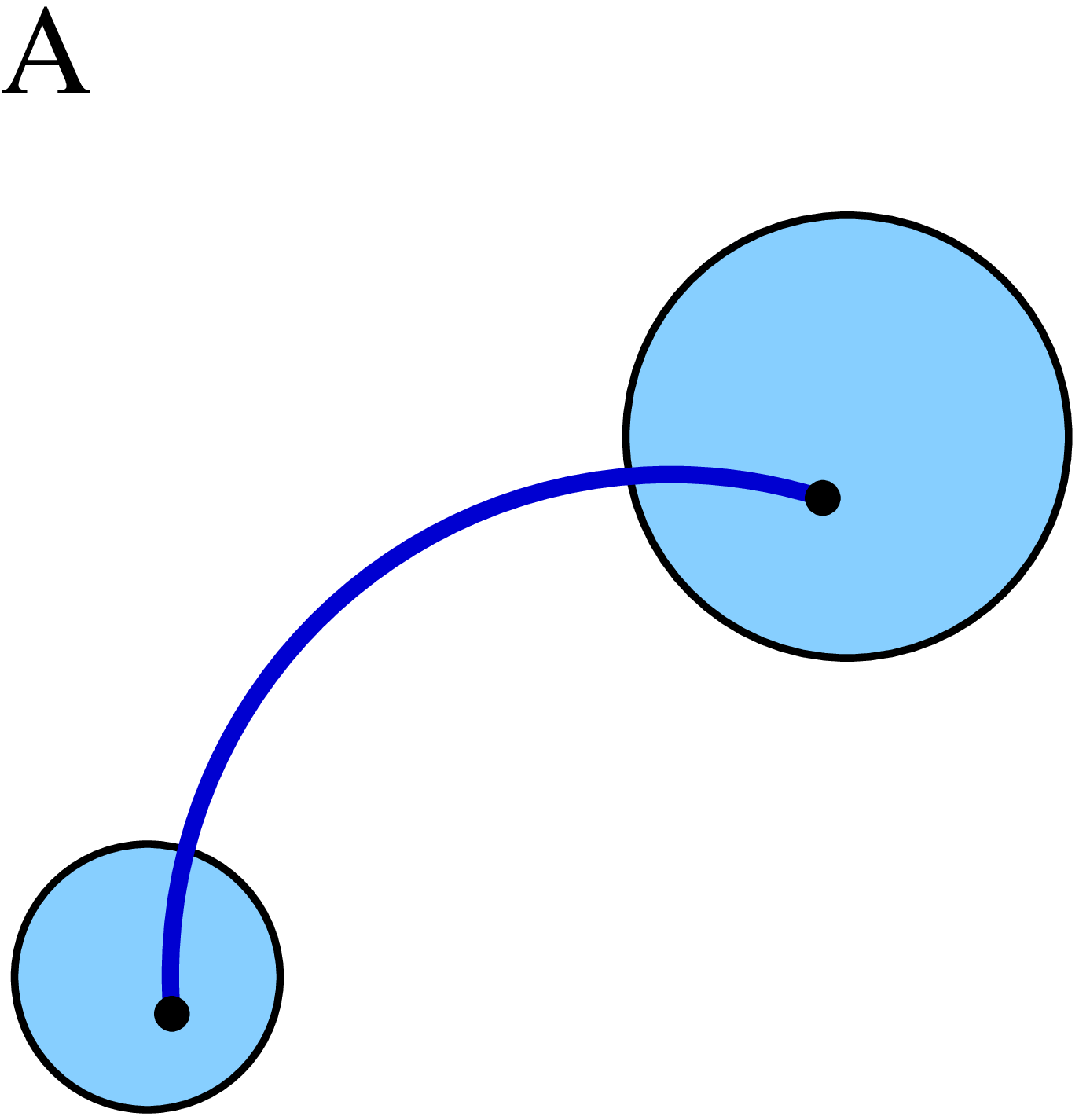}}
\scalebox{0.27}{\includegraphics[angle=0]{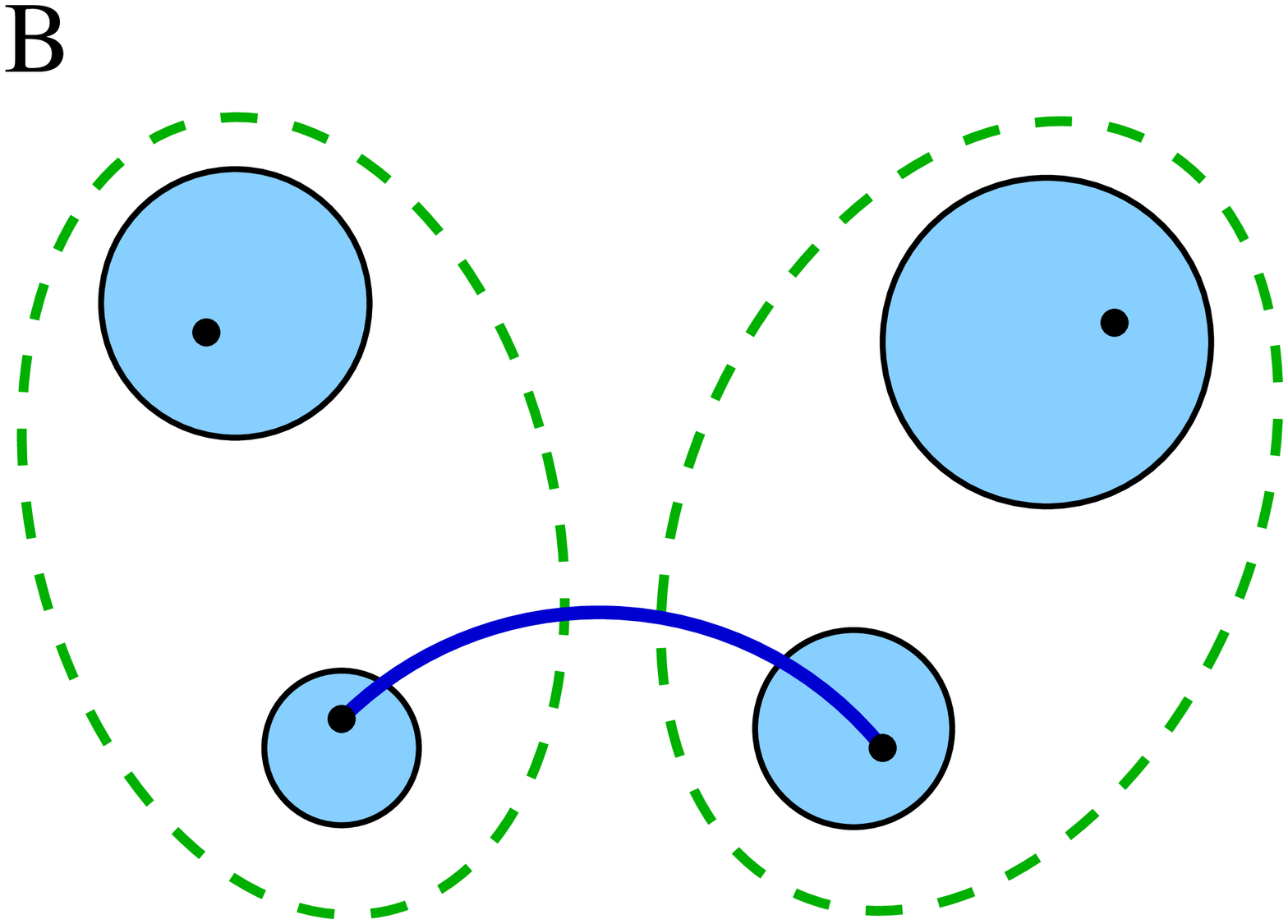}}
\\${}$\\
%%${}$\\
\scalebox{0.37}{\includegraphics[angle=270]{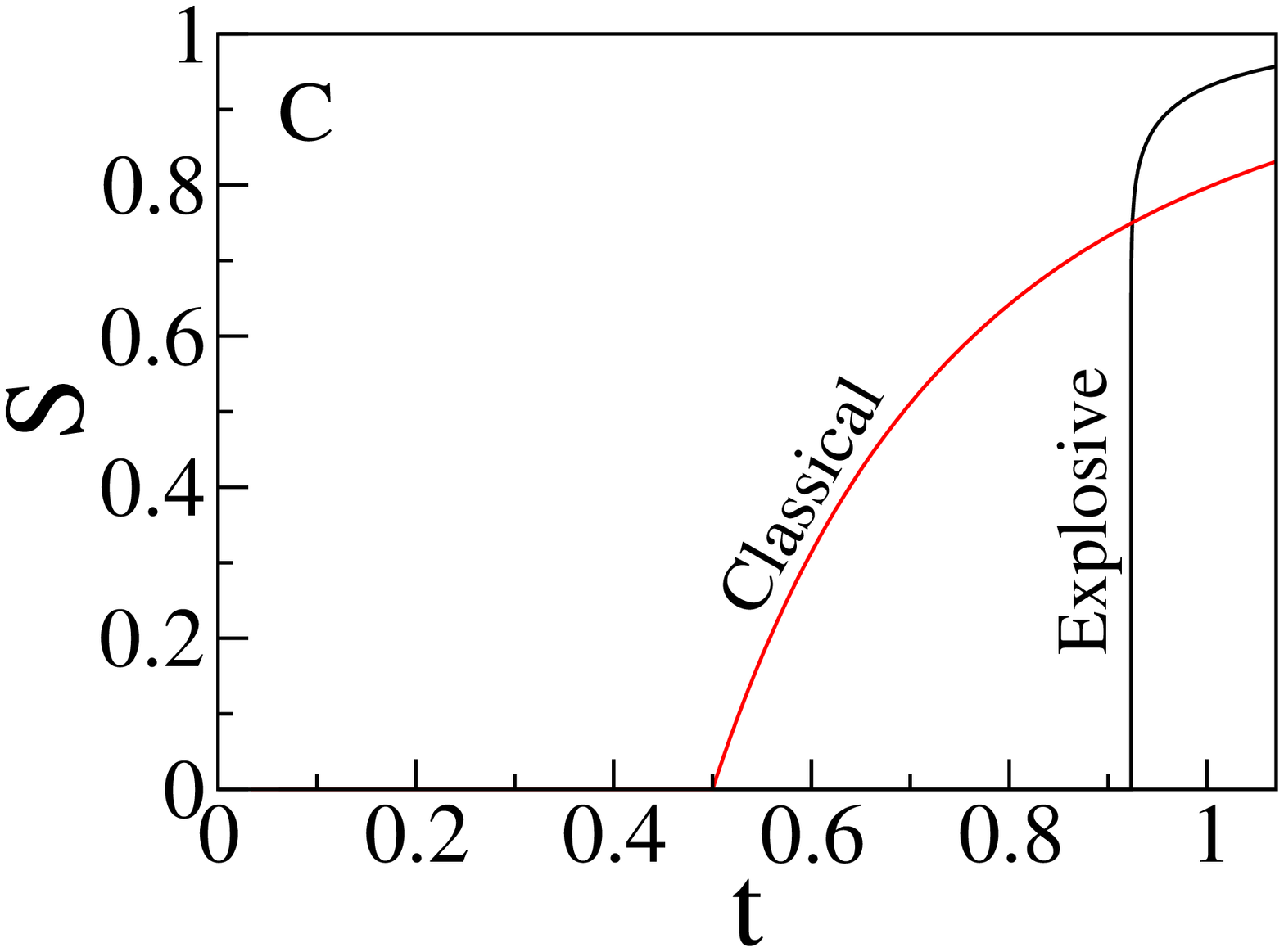}}${}\ \ \ {}$
%%\scalebox{0.27}{\includegraphics[angle=0]{figure_1C.eps}}${}\ \ {}$
%%\scalebox{0.37}{\includegraphics[angle=270]{fig_1C.eps}}
%%\scalebox{0.27}{\includegraphics[angle=0]{figure_1D.eps}}
%%\scalebox{0.37}{\includegraphics[angle=270]{fig_1D.eps}}
\scalebox{0.37}{\includegraphics[angle=270]{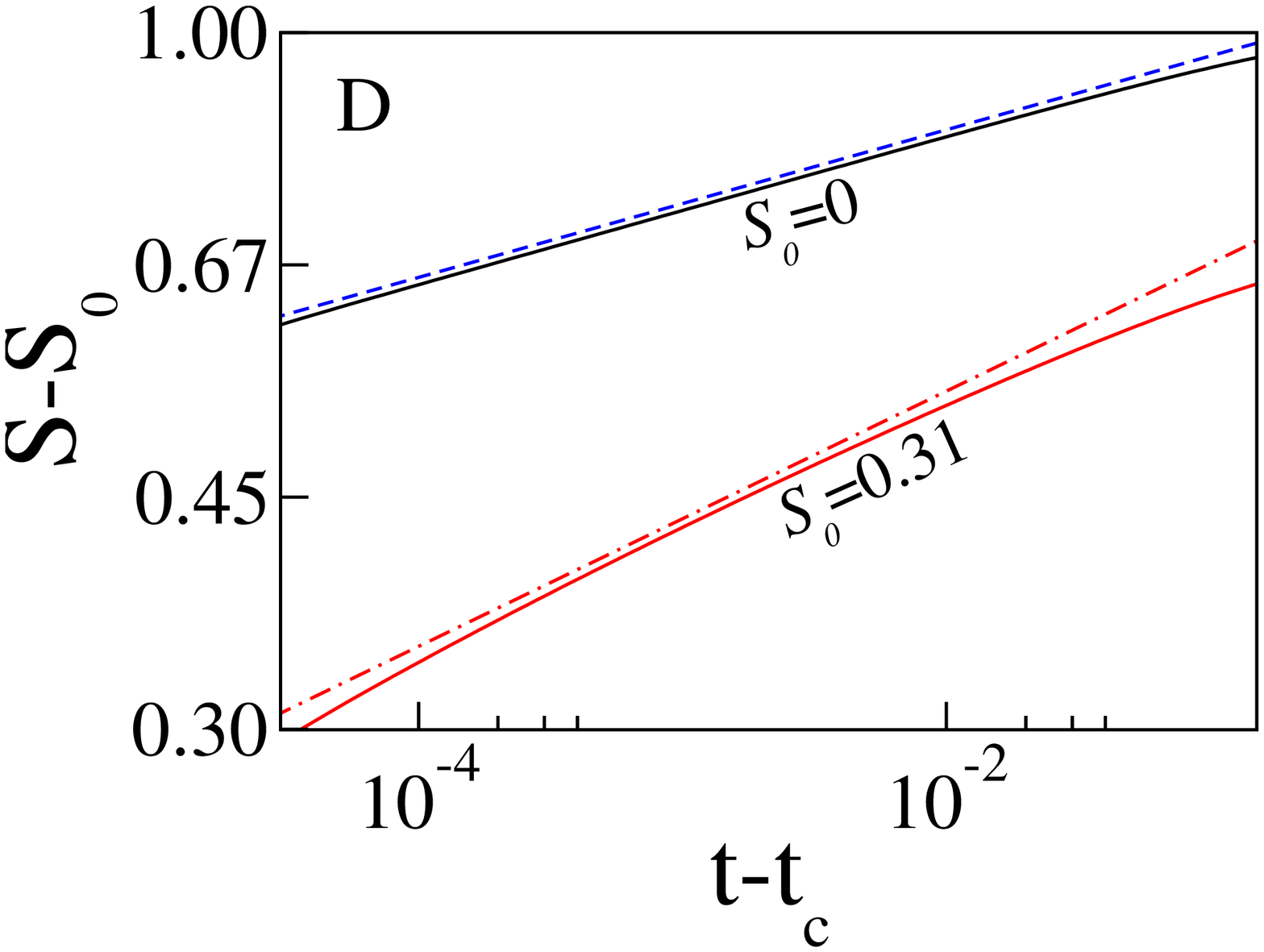}}
\end{center}
\caption{
 Percolation processes. ({\bf A}) In classical percolation, at each step, two randomly chosen nodes are connected by a new link. If these nodes belong to different clusters, these clusters merge. (B) In the ``explosive percolation'' model, at each step, two pairs of nodes are chosen at random, and for each of the pairs, the node belonging to the minimal cluster is chosen. These two nodes (and so their clusters) are connected by a new link. (C) The relative size of the percolation cluster $S$ versus $t$ (the ration of the number of links $L$ and nodes $N$) obtained from the simulation of our model with $N= 2\times10^9$ nodes ($1000$ runs). The staring configuration is $N$ bare nodes. For comparison, the corresponding result for classical percolation is shown. (D) 
Log-log plot $S$ versus $t-t_c$. The data $S$ of our simulation (black curve) is better fitted by the $a(t-t_c)^\beta$ law (dashed blue curve) than by the $S_0+b(t-t_c)^\beta$ law with $S_0=0.31$. In the latter case, $\ln(S-S_0)$ (red curve) does not follow a linear law (dash-dotted line).
} 
\label{f1}
\end{figure}

\newpage$\phantom{.}$

\begin{figure}[t]
%%%%[tbhd]
\begin{center}
%%\scalebox{0.27}{\includegraphics[angle=0]{figure_2A.eps}}\\${}$\\${}$\\ 
%%\scalebox{0.37}{\includegraphics[angle=270]{fig_2A.eps}}
\scalebox{0.37}{\includegraphics[angle=270]{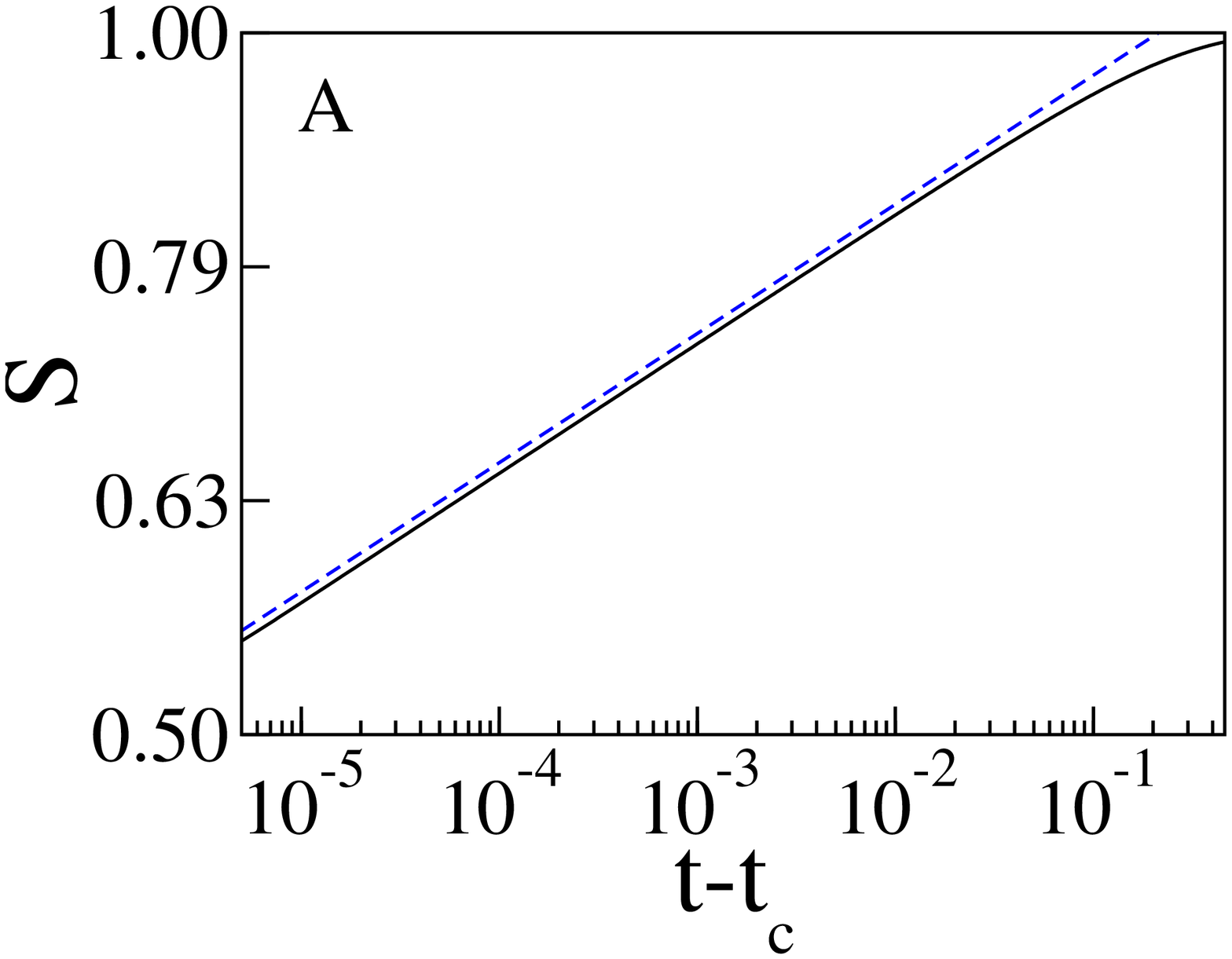}}
\\${}$\\
%%${}$\\
%%\scalebox{0.27}{\includegraphics[angle=0]{figure_2B.eps}}${}\ \ \ \ \ \ {}$
%%\scalebox{0.37}{\includegraphics[angle=270]{fig_2B.eps}}
\scalebox{0.37}{\includegraphics[angle=270]{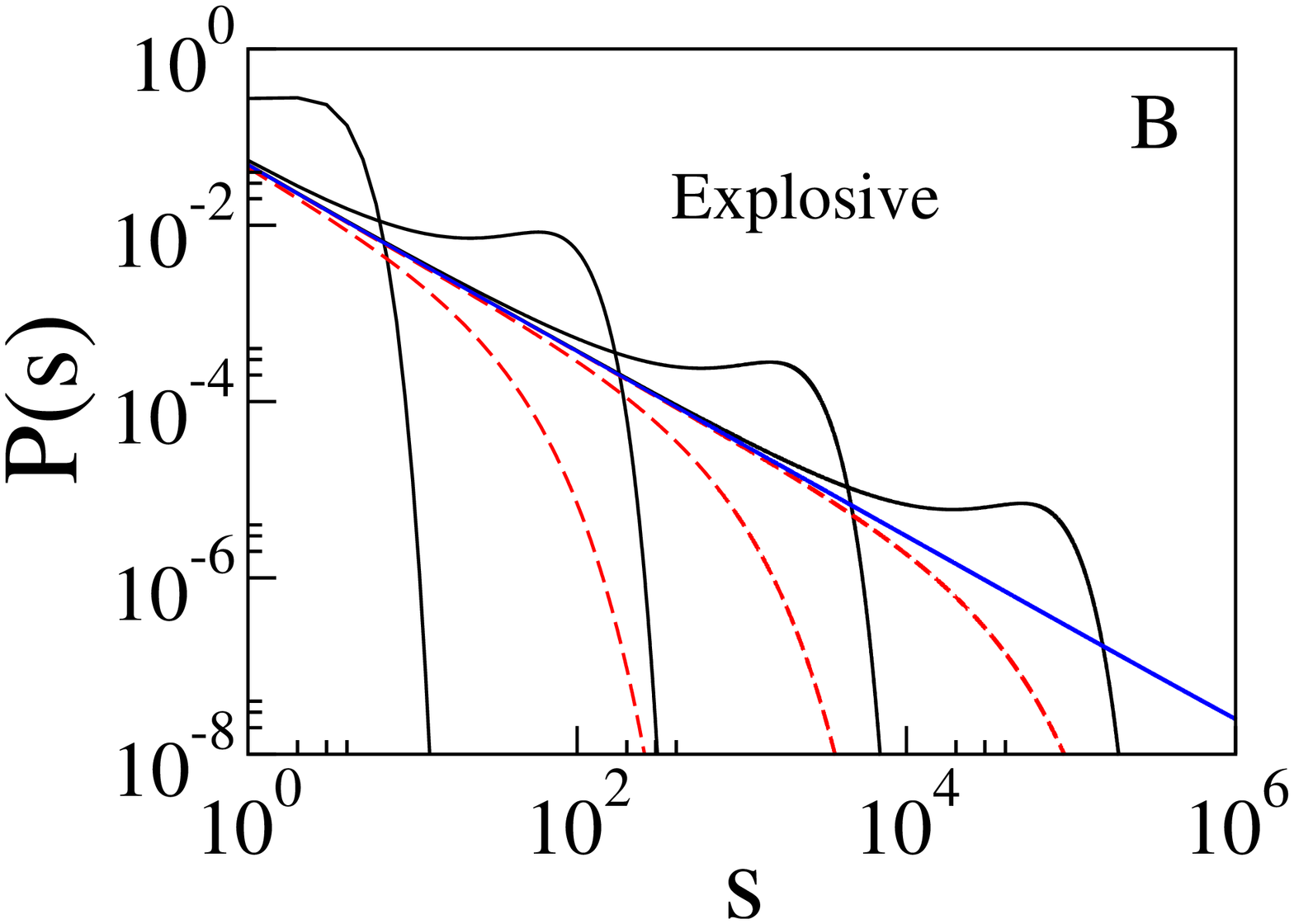}}
${}\ \ \ {}$
%%\scalebox{0.27}{\includegraphics[angle=0]{figure_2C.eps}}\\${}$\\${}$\\
%%\scalebox{0.36}{\includegraphics[angle=0]{figure_2Cnew1e5.eps}}\\${}$\\${}$\\
%%\scalebox{0.37}{\includegraphics[angle=270]{fig_2C.eps}}
\scalebox{0.37}{\includegraphics[angle=270]{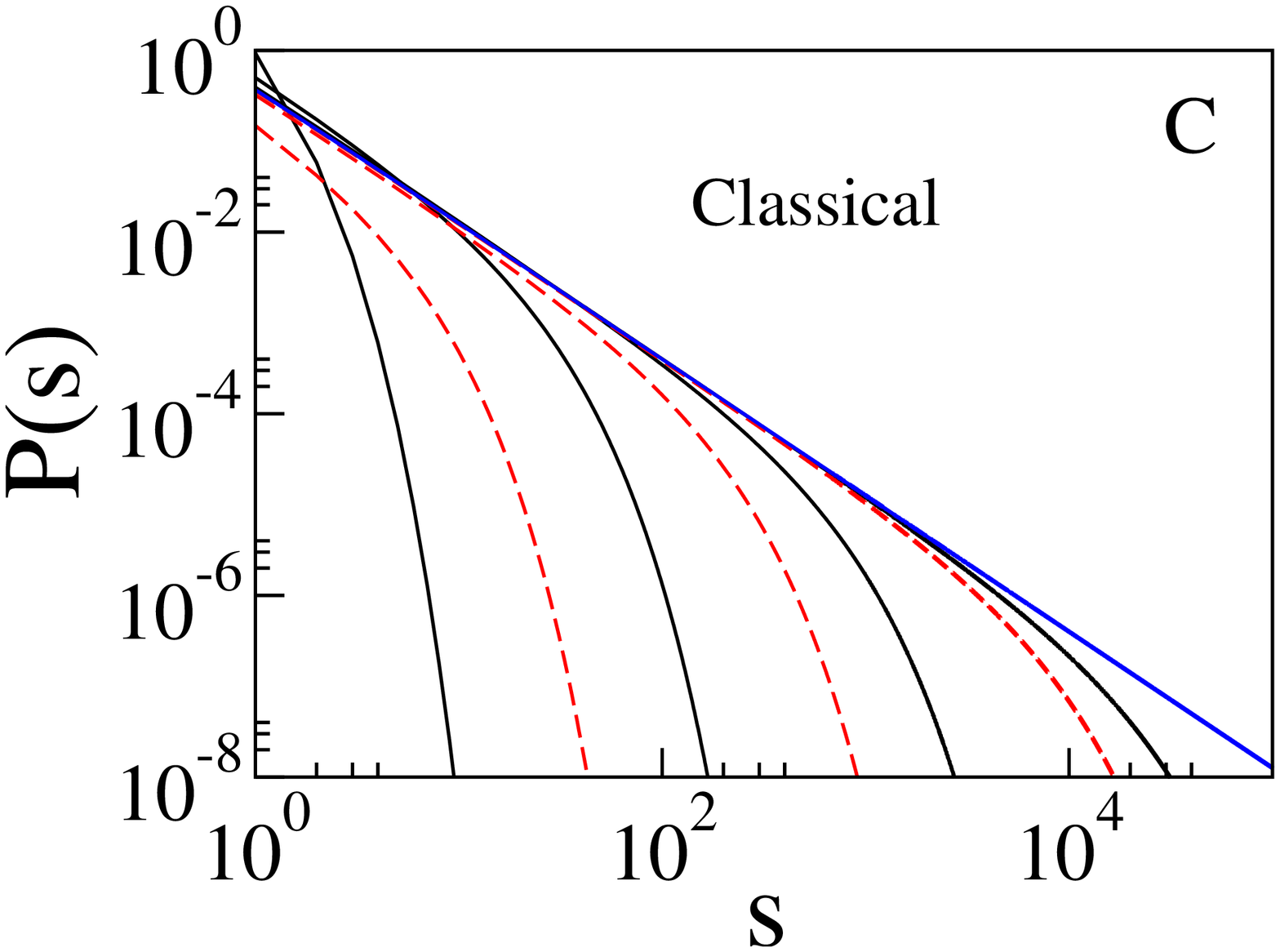}}
\\${}$\\
%%${}$\\
%%\scalebox{0.37}{\includegraphics[angle=270]{fig_2D.eps}}
\scalebox{0.37}{\includegraphics[angle=270]{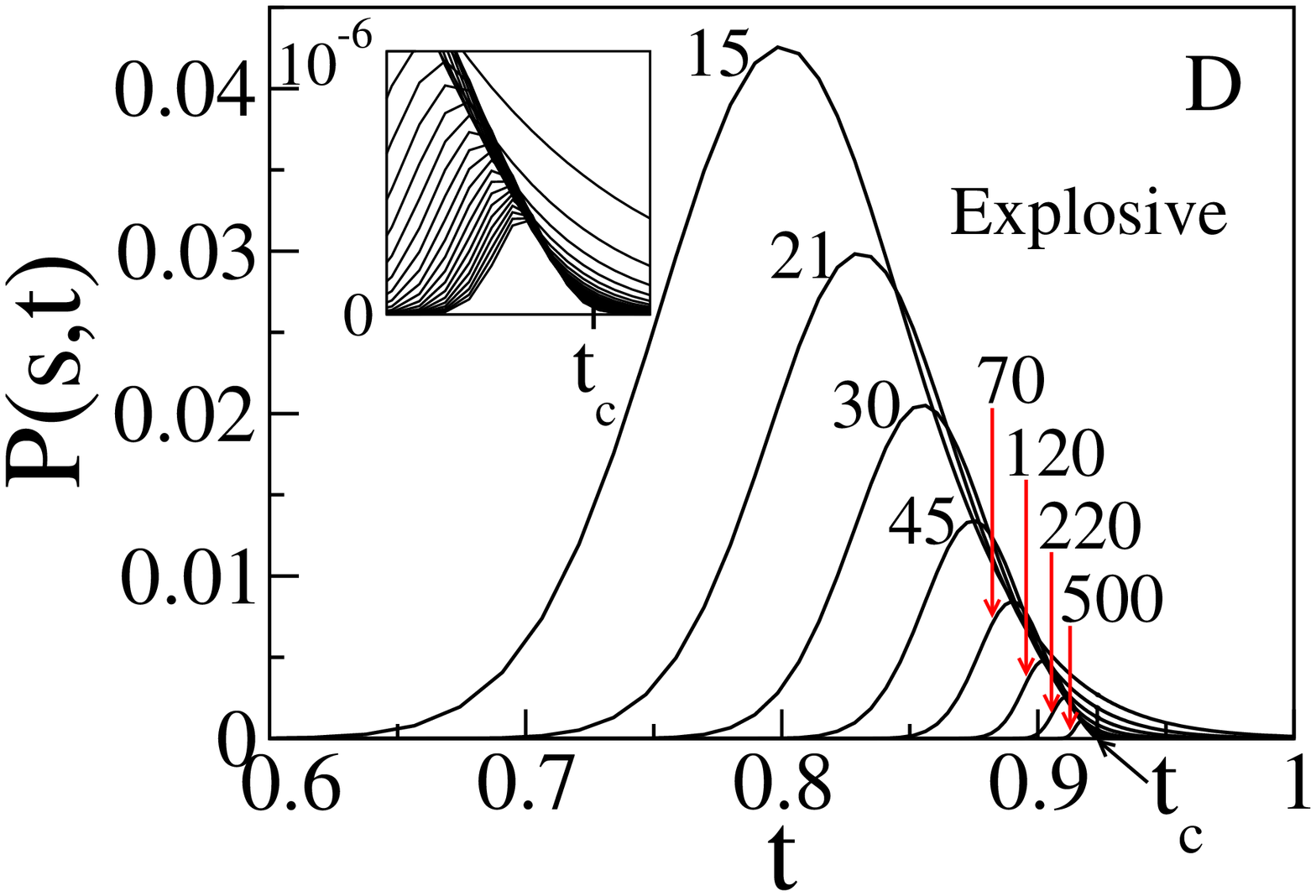}}
%%\scalebox{0.27}{\includegraphics[angle=0]{figure_2D.eps}}
${}\ \ \ \ \ \ {}$
%%\scalebox{0.27}{\includegraphics[angle=0]{figure_2E.eps}}
%%\scalebox{0.36}{\includegraphics[angle=0]{figure_2Enew1e5.eps}}
%%\scalebox{0.37}{\includegraphics[angle=270]{fig_2E.eps}}
\scalebox{0.37}{\includegraphics[angle=270]{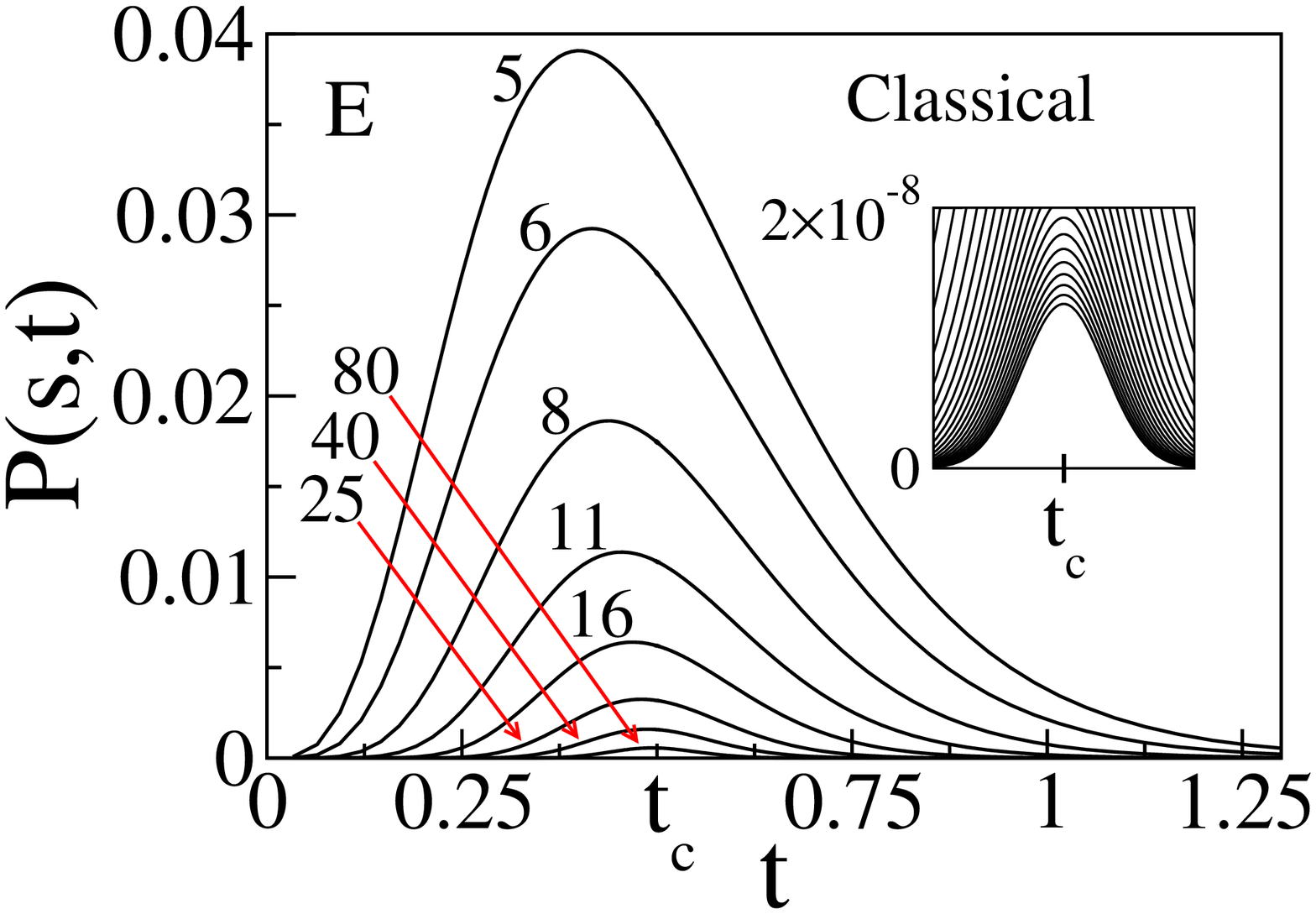}}
\end{center}
\caption{ 
Numerical solutions of the evolution equations (2) for the infinite system. 
(A) Log-log plot $S$ vs. $t-t_c$. The slope of the dashed line is $0.0555$. 
(B) The 
%%temporal 
evolution of the distribution $P(s)$ below (black lines) and above (red dashed lines) the percolation threshold for explosive percolation. The distribution at the critical point is show by the blue line. 
%%(C) The temporal evolution of $Q(s)$. 
(C) The 
%%temporal 
evolution of $P(s)$ for ordinary (classical) percolation. 
(D) Dependence of $P(s,t)$ on $t$ for a set of cluster sizes $s$ for explosive percolation. Numbers on curves indicate $s$. 
%%The inset shows 
%%(E) Dependence of $Q(s,t)$ on $t$. 
(E) $P(s,t)$ versus 
%%The same as in (D) 
for normal percolation. The insets show the $P(s,t)$ curves for large values of $s$.
} 
\label{f2}
\end{figure}

\newpage$\phantom{.}$

\begin{figure}[t]
%%%%[tbhd]
\begin{center}
%%\scalebox{0.27}{\includegraphics[angle=0]{figure_3A.eps}}
%%\scalebox{0.37}{\includegraphics[angle=270]{fig_3A.eps}}
\scalebox{0.37}{\includegraphics[angle=270]{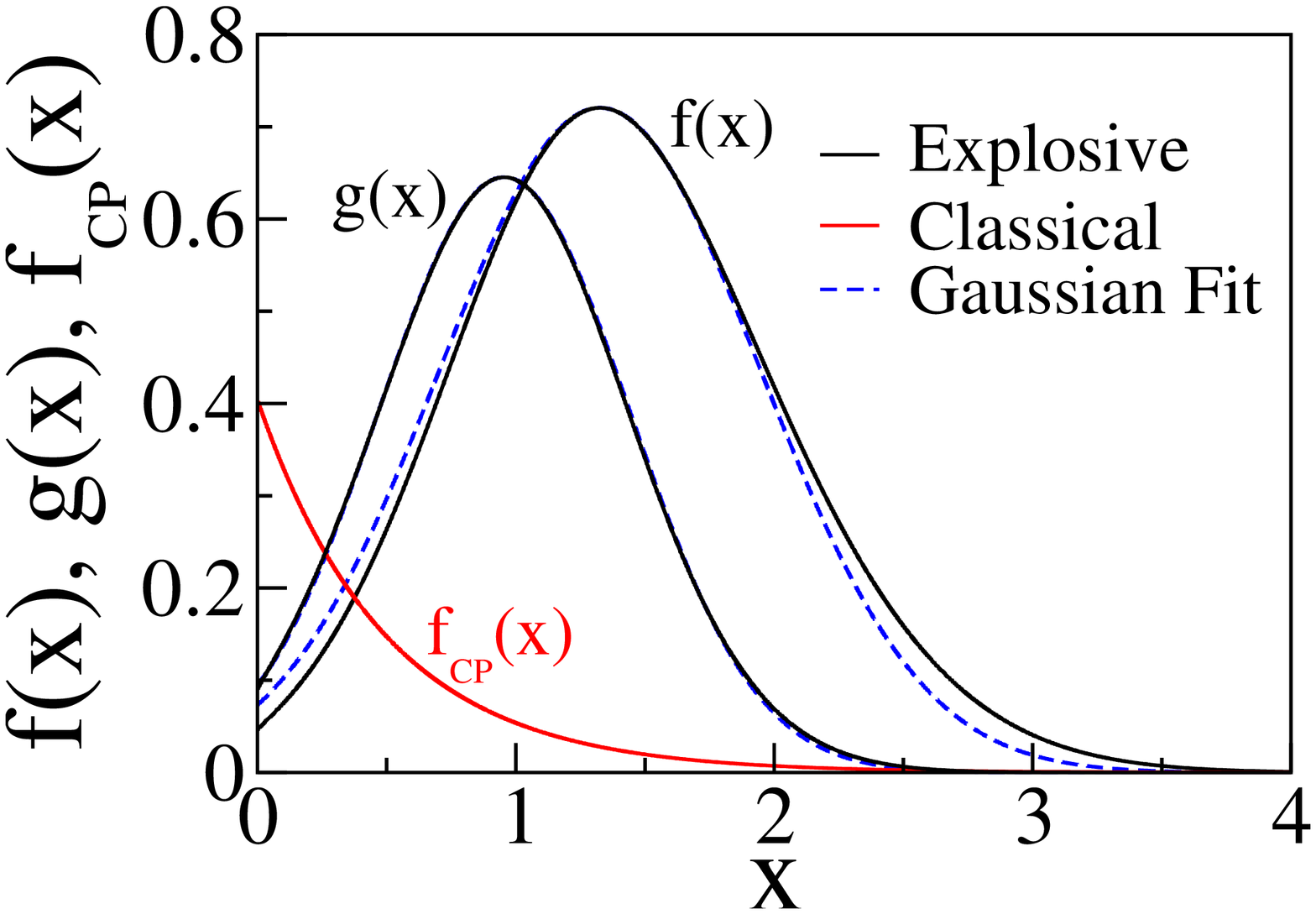}}
\end{center}
\caption{ 
Scaling functions $f(x)$ and $g(x)$ for explosive percolation ($t<t_c$) and, for comparison, the exact scaling function $f_{{\rm CP}}(x)=e^{-2x}/\sqrt{2\pi}$ for ordinary (classical) percolation. The blue dashed lines show the Gaussian fittings of the explosive percolation scaling functions.
} 
\label{f3}
\end{figure}

%%%%%%%\end{document}

\newpage

\begin{center}
{\LARGE 
``Explosive Percolation'' Transition 
\\ 
is Actually \vspace{24pt}Continuous
}
\end{center} 
%%\\
%%Supplementary Material

% Place the author information here.  Please hand-code the contact
% information and notecalls; do *not* use \footnote commands.  Let the
% author contact information appear immediately below the author names
% as shown.  We would also prefer that you don't change the type-size
% settings shown here.

\begin{center}
{\Large  
Rui A. da Costa, Sergey N. Dorogovtsev, Alexander V. Goltsev, \\
Jos\'e Fernando \vspace{30pt}F. Mendes
}
\end{center}

% Include the date command, but leave its argument blank.

%%%%%%%\date{}

%%%%%%%%%%%%%%%%% END OF PREAMBLE %%%%%%%%%%%%%%%%

%%%%%%%\begin{document} 


\baselineskip24pt


%%%%%%%\maketitle 

\begin{center}
%%{\LARGE Supplementary Material}
{\LARGE Supporting Online Material}
\end{center}


%%\begin{sciabstract}
%%  This document presents a number of hints about how to set up your
%%  {\it Science\/} paper in \LaTeX\ .  We provide a template file,
%%  \texttt{scifile.tex}, that you can use to set up the \LaTeX\ source
%%  for your article.  An example of the style is the special
%%  \texttt{\{sciabstract\}} environment used to set up the abstract you
%%  see here.
%%\end{sciabstract}

%%\noindent
%%{\Large {\bf Contents}}

\vspace{10pt}${}$
\section*{Contents}

%%1.\ \  Fitting the data for the size of the percolation cluster vs. time 
%%\\
1.\ \ Comparison between explosive percolation models
\\
2.\ \ Analysis of the evolution equation above the explosive percolation transition
\\
3.\ \ Relations between critical exponents 
\\
4.\ \ Relation between the critical time $t_c$ and exponent $\tau$

%%\begin{enumerate} 
%%
%%\item[1.\ ] Fitting the data on the size of the percolation cluster vs time. 
%%%%of simulations and numerical results for $S(t)$ 
%%
%%\item[2.\ ] Derivation of the 
%%
%%\item[3.\ ] Relations between critical exponents.
%%
%%\item[4.\ ] Relation between the critical point $t_c$ and exponent $\tau$.
%%
%%\end{enumerate}

% In setting up this template for *Science* papers, we've used both
% the \section* command and the \paragraph* command for topical
% divisions.  Which you use will of course depend on the type of paper
% you're writing.  Review Articles tend to have displayed headings, for
% which \section* is more appropriate; Research Articles, when they have
% formal topical divisions at all, tend to signal them with bold text
% that runs into the paragraph, for which \paragraph* is the right
% choice.  Either way, use the asterisk (*) modifier, as shown, to

%%\section{Fitting the data for the size of the percolation cluster vs. time} 

\newpage

Here we compare models of explosive percolation and present some details of
our analytical calculations.

\section{Comparison between explosive percolation models}

Let us show that our model provides more efficient merging of small clusters
than the model ({\it 3\/}). So if the explosive percolation transition is
continuous in our model, then the model ({\it 3\/}) also has a continuous
transition. 

Recall the selection rule in our model. 
%%In our model, at 
At each step, choose two pairs of nodes, $i$ and $j$, $k$ and
$l$, uniformly at random. Let they belong to clusters of sizes
$s_i$ and $s_j$, $s_k$, and $s_l$, respectively. Connect the smallest cluster
of the pair $s_i$ and $s_j$, with the smallest cluster of the pair $s_k$, and
$s_l$. 
One can immediately see that this rule is equivalent to the following one. Let
$f(s,s')$ be an arbitrary monotonously growing function of its arguments,
e.g., $f(s,s')=ss'$. Consider four
possibilities to add a new link, $ik$, $il$, $jk$, and $jl$, and choose that
one which provides the smallest value of $f(s_i,s_k)$, $f(s_i,s_l)$,
$f(s_j,s_k)$, and $f(s_j,s_l)$, see Fig.~4A in the Supporting online material.  

\begin{figure}[b]
%%%%[tbhd]
\begin{center}
\scalebox{0.25}{\includegraphics[angle=0]{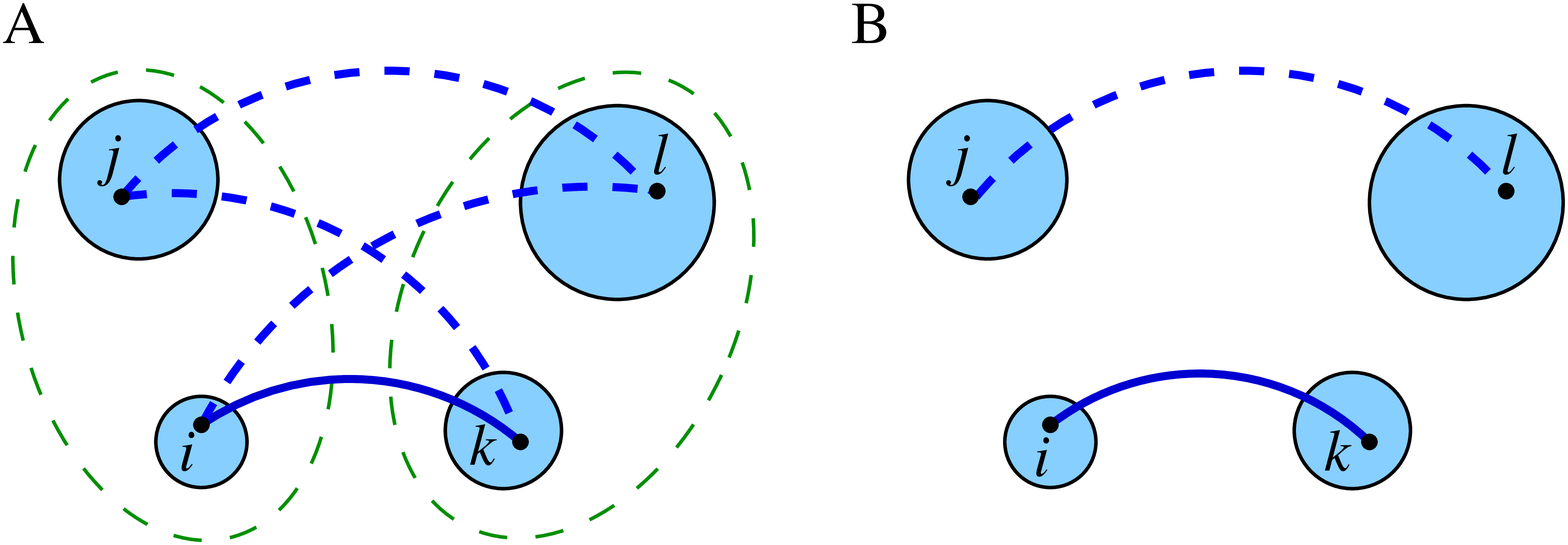}}
\end{center}
\caption{ 
Comparison of the linking rules in our model (A) and in the model ({\it
  3\/}) (B).  
} 
\label{fs1}
\end{figure}

%%In the model ({\it 3\/}) (product selection rule), at each step, one chooses
%%uniformly at random nodes $i$ $j$, $k$ and $l$, and selects from 
In the model ({\it 3\/}), the product selection rule is as follows. At each
step choose nodes $i$ $j$, $k$ and $l$ uniformly at random, and select from 
%%only 
two
possibilities, either connect $i$ and $k$ or connect $j$ and $l$, choosing the pair
with the smallest of the products $s_is_k$ and $s_js_l$, see Fig.~1B. 

Then the only difference between 
our model and the model ({\it 3\/}) is that we select from four possibilities 
(comparison of $s_is_k$, $s_is_l$, $s_js_k$ and $s_js_l$) while in the model 
({\it 3\/}) the selection is from only two possibilities 
(comparison of $s_is_k$ and $s_js_l$). 
%%the model ({\it 3\/}) and our model is that
%%in the former the selection is from only two possibilities
%%(comparison of $s_is_k$ and $s_js_l$) while we select from four possibilities
%%(comparison of $s_is_k$, $s_is_l$, $s_js_k$ and $s_js_l$). 
Therefore our
choice guarantees merging of clusters with the product of sizes $ss'$ which is equal or
smaller than in the model ({\it 3\/}). 
So our process should generate discontinuity even more efficiently
than in the model ({\it 3\/}). Consequently, if our model has a continuous
phase transition, that the model ({\it 3\/}) also must have a continuous
transition.

\section{Analysis of the evolution equation above the explosive percolation
  transition}

Let us consider the master equation for our model 
\begin{equation}
\frac{\partial P(s,t)}{\partial t}
= s \sum_{u+v=s} Q(u,t)Q(v,t) -2 sQ(s,t)
, 
\label{e10}
\end{equation}
where 
\begin{equation}
Q(s) = [P_{\rm cum}(s)
%%+S
+P_{\rm cum}(s+1)+2S]P(s) 
= [2-2P(1)
%%-2P(2)
-\ldots-2P(s-1)-P(s)]P(s)
. 
\label{e20}
\end{equation}
(Here we only consider the case of $m=2$.)  
We introduce generating functions for the distributions $P(s)$ and $Q(s)$, 
\begin{equation}
\rho(z) \equiv \sum_{s=1}^\infty P(s) z^s 
\label{e30}
\end{equation}
and 
\begin{equation}
\sigma(z) \equiv \sum_{s=1}^\infty Q(s) z^s
.
\label{e40}
\end{equation}
Note that $\rho(z=1)=1-S$ and $\sigma(z=1)=1-S^2$, where $S$ is the relative size of the percolation cluster. 
Using these generating functions for the distributions we represent the master equation (\ref{e10}) in the following form:
\begin{equation}
%%\partial_t
\frac{\partial
%%[1-\rho(z,t)]
}{\partial t}
[1-\rho(z,t)] 
= -
%%\partial_{\ln z}
\frac{\partial}{\partial \ln z}
[1-\sigma(z,t)]^2
.
\label{e50}
\end{equation}
In ordinary percolation, $\sigma(z,t)$ in this equation is substituted by $\rho(z,t)$, namely
\begin{equation}
%%\partial_t
\frac{\partial\rho(z,t)
}{\partial t} 
= 
2[\rho(z,t)-1]
%%\partial_{\ln z}
\frac{\partial\rho(z,t)}{\partial \ln z}
.
\label{e60}
\end{equation}
In the critical region above the percolation threshold, Eq.~(\ref{e20}) gives  
%%asymptotically 
%%
\begin{equation}
Q(s) \cong 2S P(s)
%% .
\label{e70}
\end{equation}
asymptotically at large $s$, which leads to a simple relation between the generating functions $\sigma(z)$ and $\rho(z)$ in the range $z$ close to $1$. Indeed, 
\begin{equation}
1-S^2-\sigma(z)= \sum_s Q(s)[1-z^s] \cong \sum_s 2SP(s)[1-z^s] = 2S[1-S-\rho(z)]
, 
\label{e80}
\end{equation}
so
\begin{equation}
1-\sigma(z) \cong 2S[1-\rho(z)-S/2]
%% .
\label{e90}
\end{equation}
if $z$ is close to $1$ in the critical region at $t>t_c$. 
One can check this relation at $z=1$, namely,  
%%The relative size of the percolation cluster, $S$, is 
%%
\begin{equation}
1-\sigma(1) = S^2 = 2S[1-\rho(1)-S/2] = 2S(S-S/2)
.
\label{e100}
\end{equation}
Therefore, in the critical region above the percolation threshold, at $z$ close to $1$, the master equation takes a convenient form, 
\begin{equation}
%%\partial_t
\frac{\partial\rho(z,t)
}{\partial t} 
= 
8S^2(t)[\rho(z,t)-1+S(t)/2]
%%\partial_{\ln z}
\frac{\partial\rho(z,t)}{\partial \ln z}
.
\label{e110}
\end{equation}
Note that this equation essentially differs from Eq.~(\ref{e60}) for ordinary percolation because of the terms $S(t)$ on the right-hand side. Nonetheless Eq.~(\ref{e110}) can be analysed in the same way as for ordinary percolation. 

Our numerical solution of Eq.~(\ref{e10})  
%%enable us to assume 
showed convincingly that at the critical point, the distribution $P(s,t_c)$ is power-law, namely, at large $s$, $P(s,t_c)\cong f(0)s^{1-\tau}$. Here $f(0)$ is the critical amplitude for this distribution. This is also the value of the scaling function for this distribution, $f(x=0)$, see below. 
%%Let us assume that 
Let us show that, 
if at the critical point $P(s,t_c) \propto s^{1-\tau}$, then Eq.~(\ref{e110}) has a solution with $1-\rho(z=1,t)=S(t) \propto (t-t_c)^\beta$ in the critical region, which just means that the transition is continuous. 
%%the assumptions of $P(s,t_c) \propto s^{1-\tau}$ and of $S \propto (t-t_c)^\beta$ for the master equation (\ref{e110}) are self-consistent, that is the transition is continuous. 
The existence of this solution can be demonstrated in the following way. 
Let us assume that $P(s,t_c)\cong f(0)s^{1-\tau}$ at large $s$, and $S \cong B(t-t_c)^\beta$ at small $t-t_c$ and check whether this assumption is correct or not. Here we assume that $f(0)$ and the exponent $\tau$ are known (numerical solution gave $f(0)=0.04618(2)$ and $\tau=2.04762(2)$), while $B$ and the exponent $\beta$ are to be found. 
%%If we obtain reasonable $B$ and $\beta$, then this solution exists. 

We use the power-law asymptotics of the distribution $P(s,t_c)\cong f(0)s^{1-\tau}$ as the initial condition for Eq.~(\ref{e110}). This corresponds to the following singularity of the generating function at $z=1$: 
\begin{equation}
1-\rho(z,t_c) = {\rm analytic\ terms} - f(0)\Gamma(2-\tau)(1-z)^{\tau-2}
. 
\label{e120}
\end{equation}
Introducing $\epsilon\equiv(t-t_c)^{2\beta+1}$ and $x\equiv\ln z$, we transform Eq.~(\ref{e110}) to the following form:  
\begin{equation}
\frac{\partial\rho}{\partial\epsilon} = \frac{8B^2}{1+2\beta}\Bigg(\rho-1+\frac{B}{2}\epsilon^{\beta/(1+2\beta)}\Bigg)\frac{\partial\rho}{\partial x}
. 
\label{e130}
\end{equation}
To solve this equation, we use the hodograph transformation approach. We pass from $\rho=\rho(x,\epsilon)$ to $x=x(\rho,\epsilon)$, which leads to a simple linear partial differential equation for $x(\rho,\epsilon)$ and enables us to find the general solution 
\begin{equation}
%%\frac{\partial x}{\partial\epsilon} = 
\ln z = \frac{8B^2}{1+2\beta}\Bigg[1-\rho-\frac{B}{2}\frac{\epsilon^{\beta/(1+2\beta)}}{1+\beta/(1+2\beta)}\Bigg]\epsilon + F(\rho)
, 
\label{e140}
\end{equation}
where the function $F(\rho)$ is obtained from the initial condition (\ref{e120}), which gives the solution  
%%Taking into account the initial condition (\ref{e120}) gives the solution 
%%
\begin{equation} 
\ln z = \frac{8B^2}{1{+}2\beta}\Bigg[1-\rho-\frac{B}{2}\frac{(t-t_c)^\beta}{1{+}\beta/(1{+}2\beta)}\Bigg](t-t_c)^{1+2\beta} -[f(0)]^{-1/(\tau-2)}|\Gamma(2-\tau)|^{-1/(\tau-2)}[1-\rho]^{1/(\tau-2)}
. 
\label{e150}
\end{equation}
We set $z=1$, and taking into account 
%%At $z=1$, we have 
$1-\rho(1)=S=B(t-t_c)^\beta$ 
%%in this relation. Comparing 
and comparing resulting powers and coefficients in Eq.~(\ref{e150}), we obtain a relation between the critical exponents, 
\begin{equation}
\tau=2+\frac{\beta}{1+3\beta}
,  
\label{e160}
\end{equation}
and express the critical amplitude $B$ for the relative size of the percolation cluster in terms of the critical amplitude $f(0)$ for the distribution $P(s,t_c)$, 
\begin{equation}
B = \Bigg[4\frac{(\tau-1)(7-3\tau)}{3-\tau}\Bigg]^{(\tau-2)/(7-3\tau)}[f(0)]^{1/(7-3\tau)} |\Gamma(2-\tau)|^{1/(7-3\tau)}
.  
\label{e170}
\end{equation}
%% 
%%The relations (\ref{e160}) and (\ref{e170}) show that the transition is continuous.  
The relation (\ref{e160}) precisely agrees with our numerical results, $\tau=2.04762(2)$ and $\beta=0.0555(1)$. Substituting $f(0)=0.04618(2)$ (our numerical result) into Eq.~(\ref{e170}) gives $B=1.075$, which agrees with the corresponding value $1.080$ obtained by solving the muster equation numerically. 

%%In this report we do not present a strict derivation of the power-law distribution at the critical point. Let us outline this program. 
Furthermore, the power-law distribution at the critical point can be justified strictly by using an equation for scaling functions. In the normal phase ($t<t_c$), we derived an equation for the scaling functions in this problem. This is a nonlinear integral-differential eigenfunction equation, 
%%a nonlinear eigenfunction equation for the scaling functions in this problem. This is a nonlinear  integral-differential equation, 
where eigenfunctions are the scaling functions for $P(s,t)$ and $Q(s,t)$, see Eq.~(\ref{e180}), and a critical exponent, e.g., $\tau$, plays the role of an eigenvalue. This equation can be solved numerically, which is, however, a difficult task. We verified that the scaling functions and $\tau$, which we found numerically by solving Eq.~(\ref{e10}), satisfy this equation. This shows that the distributions are power-law at the critical point.

\section{Relations between critical exponents} 

Here we present a list of relations for critical exponents for explosive percolation ($m=2$). 

In the critical region, the distributions $P(s,t)$ and $Q(s,t)$ have a scaling form:
\begin{eqnarray}
&&
P(s,t) = s^{1-\tau} f(s\delta^{1/\sigma})
\nonumber
\\[5pt]
&&
Q(s,t) = s^{3-2\tau} g(s\delta^{1/\sigma})
,  
\label{e180}
\end{eqnarray}
where $\delta=|t-t_c|$. Note that the critical exponents below and above the transition are equal, while the scaling functions below $t_c$ differ dramatically from those above the transition. The exponent $\beta$ of the size of the percolation cluster is expressed in terms of $\tau$ and $\sigma$ as follows: 
\begin{equation}
\beta=\frac{\tau-2}{\sigma}
.  
\label{e190}
\end{equation}
For the first moments of the distributions $P(s)$ and $Q(s)$, critical exponents are $\gamma_{\scriptscriptstyle P}$ and $\gamma_{\scriptscriptstyle Q}$, respectively. Namely, $\langle s \rangle_{\scriptscriptstyle P}\propto |\delta|^{-\gamma_{\scriptscriptstyle P}}$ and $\langle s \rangle_{\scriptscriptstyle Q}\propto |\delta|^{-\gamma_{\scriptscriptstyle P}}$. These exponents are expressed in therms of $\tau$ and $\sigma$ as follows:  
\begin{eqnarray}
&&
\gamma_{\scriptscriptstyle P}=\frac{3-\tau}{\sigma}
,
%%\nonumber
\\[5pt]
&&
\gamma_{\scriptscriptstyle Q}=\frac{5-2\tau}{\sigma}
.  
\label{e200}
\end{eqnarray}
Note the relation between $\gamma_{\scriptscriptstyle P}$ and $\gamma_{\scriptscriptstyle Q}$:  
\begin{equation}
\gamma_{\scriptscriptstyle P} + 1 = 2\gamma_{\scriptscriptstyle Q} 
.  
\label{e210}
\end{equation}

Finally, we give the full set of critical exponents, the fractal dimension $d_f=2/\sigma$ and the upper critical dimension $d_u = d_f+2\beta$ in terms of the exponent $\beta$ in the case of $m=2$: 
\begin{eqnarray}
&&
\tau = 2 + \frac{\beta}{1+3\beta}
,
%%\nonumber
\\[5pt]
&&
\sigma = \frac{1}{1+3\beta}
,
%%\nonumber
\\[5pt]
&&
\gamma_{\scriptscriptstyle P} = 1+2\beta
,
%%\nonumber
\\[5pt]
&&
\gamma_{\scriptscriptstyle Q} = 1+\beta
,
%%\nonumber
\\[5pt]
&&
d_f = 2(1+3\beta)
,
%%\nonumber
\\[5pt]
&&
d_u = 2(1+4\beta)
.  
\label{e220}
\end{eqnarray}

\section{Relation between the critical time $t_c$ and exponent $\tau$}

We can obtain approximate relations between $t_c$ and $\tau$ or between the critical amplitude $f(0)$ and $\tau$ by applying the sum rule $\sum_s P(s) =1$ at the critical point. 
Two estimates are possible. 
One can estimate $P(s,t_c)$ by its asymptotics $f(0)s^{1-\tau}$, which gives 
\begin{equation}
f(0) \zeta(\tau-1) \approx 1
,   
\label{e290}
\end{equation}
where $\zeta(x)=\sum_{s=1}^\infty s^{-x}$ is the Riemann zeta function. If $\tau$ is close to $2$, then $\zeta(\tau-1)\cong 1/(\tau-2)$, so we have $\tau-2 \approx f(0)$. This estimate shows that the small values of $\tau-2$ and $f(0)$ are interrelated. Recall that we obtained numerically $\tau=2.04762(2)$ and $f(0)=0.04618(2)$. 

In the second estimate we use the following approximation: $P(s,t_c) \approx P(1,t_c)s^{1-\tau}$. We find $P(1,t)$ explicitly by solving the master equation (\ref{e10}), which gives
\begin{equation}
P(1,t) = \frac{2}{1+e^{4t}}
,   
\label{e300}
\end{equation}
so we have 
\begin{equation}
\frac{2}{1+e^{4t_c}}\zeta(\tau-1) \approx 1
.   
\label{e310}
\end{equation}
Using $t_c=0.923207508(2)$ obtained numerically, we find approximately $\tau-2\approx 0.05$, that is, the exponent $\tau$ is close to $2$ when $t_c$ is close to $1$. 
%%, the exponent $\tau$ is close to $2$. 
%%which shows that the small values of $\tau-2$ and $1-t_c$ are interrelated.  

\end{document}